\documentclass[preprint,showpacs,preprintnumbers,amsmath,amssymb]{revtex4-2}
\usepackage{bm}
\usepackage{amsmath}
\usepackage{amssymb}
\usepackage{amsfonts}
\usepackage{graphicx}
\begin{document}

\title{ \textbf{
    Teleportation in Proton Systems Revisited
} }

\author{H.~Wita{\l}a}
\affiliation{M. Smoluchowski Institute of Physics, 
Faculty of Physics, Astronomy and Applied Computer Science,
Jagiellonian University, PL-30348 Krak\'ow, Poland 
}

\date{\today}

\begin{abstract}
  We discuss the peculiarities of scattering in a three-proton system induced
  by the presence of an entangled proton–proton Bell-like state in the initial
  configuration. We formulate the problem using the standard spin formalism
  of nuclear physics in order to follow in detail the reaction pathways
  leading to the teleportation of a quantum-mechanical state within this system.

  By performing numerical simulations of the system and calculations based
  on a realistic nucleon–nucleon potential, we evaluate the relevant spin
  observables and search for a simple, experimentally feasible signal
  that could provide clear evidence of quantum teleportation.

  For the case in which proton $2$, belonging to the entangled proton pair
  $23$, interacts with hydrogen target $1$, thereby triggering the
  teleportation process, we find clear evidence of teleportation through
  measurements of the final polarization of proton $3'$, even for small values
  of the polarization of target proton $1$. The dominant component of this
  polarization is transferred almost entirely from target proton $1$ to proton
  $3'$ in the angular region characterized by strong entanglement.
  
  For an unpolarized target proton $1$, quantum teleportation is no longer
  possible. The only unambiguous signature of the residual quantum
  correlations would be provided by measurements of the spin correlations
  within the strongly entangled final-state proton pair $1'2'$. Indirect
  evidence may also be obtained from measurements of the polarization
  of proton $2'$ in the final state, although the corresponding effect
  is relatively weak.
\end{abstract}


\maketitle \setcounter{page}{1}

\section{Introduction}
\label{intro}

Recent studies have shown that entangled Bell-like proton–proton pairs
can be produced in proton–proton (pp) elastic scattering using an unpolarized
beam and target \cite{shen_2025,wit_unp_pd}. Moreover, the exclusive breakup
reaction of unpolarized proton–deuteron (pd) scattering can also provide
high-quality Bell-like states of two protons in kinematically complete
quasi-free-scattering (QFS) and final-state-interaction (FSI)
configurations \cite{wit_unp_pd}. The availability of high-intensity
entangled pp states has made it possible to seriously consider an
experimental verification of the intriguing possibility of teleporting
a quantum-mechanical state between two protons within a three-proton
system \cite{wit_unp_pd}.

Such an experiment, proposed in \cite{teleport} and discussed 
in \cite{wit_unp_pd}, is essentially based on two elements. The first
is the formation of entangled pp pairs, and the second is the scattering
of one of the entangled protons off a polarized proton target. The latter
process is responsible for the teleportation of the quantum state of
the target proton to the second entangled proton, leaving the two
remaining protons in a new entangled state.

For this experiment to be successful, in addition to the requirement
of producing a specific Bell-type entangled state of two protons, the
actual occurrence of teleportation requires a specific single-term
structure of the pp scattering transition matrix, with the dominant
contribution originating from one particular Bell component.

However, the proposed experimental setup, which relies on such entangled
states, requires a polarized hydrogen target, currently representing an
insurmountable obstacle to its practical implementation.

In Ref.~\cite{wit_unp_pd}, we proposed an alternative approach based
exclusively on unpolarized states, thereby completely eliminating the need
for polarized ones. However, when only an unpolarized hydrogen target
is employed, the teleportation process itself can no longer be directly
investigated, and the analysis is restricted to the residual spin
correlations. As a signature of these residual correlations, we proposed
measuring the strong spin correlation between the two entangled final-state
protons.

In the present work, we investigate additional signatures of spin correlations
and illustrate quantum-state teleportation by numerically simulating the
proposed experimental setup using a realistic nucleon–nucleon (NN) interaction.

We demonstrate the presence of strong teleportation signatures when a
polarized hydrogen target is employed, over a broad range of
target polarizations.

To gain a deeper understanding of teleportation in a three-proton system,
we formulate the problem within the standard spin formalism commonly used
in nuclear physics. This approach is essential for identifying signatures
of teleportation based solely on measurements of the final polarizations
of the outgoing protons at low energies, without requiring the considerably
more challenging determination of spin correlations in the newly formed
entangled state.

This formalism is also crucial for understanding that the formation of
the new entangled final state is not a consequence of the teleportation
of the strongly entangled Bell state produced in the first scattering.
Rather, it arises directly from the dominance of a single Bell-state
contribution in the transition matrix governing the second scattering.

Restricting the experiment to an unpolarized hydrogen target effectively
switches off the teleportation process, leaving only the residual spin
correlations available for investigation. Our aim is therefore to identify
experimentally accessible observables that would allow these correlations
to be verified without requiring a measurement of the large spin
correlation between the two entangled final-state protons.
We show that the final polarization of proton $2'$ in the entangled state
produced in the second scattering, when compared with the corresponding
polarization obtained in the absence of the unpolarized hydrogen target,
can serve as such an observable.

We also investigate and discuss other peculiarities that arise when
working with entangled states of protons at low energies around $10$~MeV.
In this energy region, consecutive proton scatterings reduce the energies
of the outgoing protons and improve the conditions for entanglement
formation, thereby providing numerous pairs of entangled protons.
By considering two such entangled pairs and allowing one proton
from each pair to interact, entanglement transfer can occur.
As a result, the final entanglement is established between protons
originating from different pairs.

In Sec.~\ref{formNN}, for the convenience of the reader, the basics
of the scattering formalism for beam and target states prepared in
specific spin configurations, together with the main results of
Ref.~\cite{wit_unp_pd}, are briefly outlined. The results of the numerical
simulations of the teleportation experiment are presented and discussed
in Subsec.~\ref{simul}, while the peculiarities arising in systems
containing multiple entangled proton pairs are discussed in
Subsec.~\ref{transfer}. Section~\ref{sumary} contains a summary and conclusions.

\section{Proton Scattering with Entangled Initial States}
\label{formNN}

In what follows, we investigate the role of entanglement in the initial
configuration of a three-proton system and its effect on the final
scattering state.
A representative example of maximal entanglement is given by
the Bell-state basis \cite{bookqinf}, defined as:
\begin{eqnarray}
  | \psi_1\rangle  &\equiv& | \Phi^+ \rangle = \frac {1} {\sqrt{2}}
  (| +\frac {1} {2} +\frac {1} {2}\rangle  +
   | -\frac {1} {2} -\frac {1} {2}\rangle ) \equiv
 \frac {1} {\sqrt{2}} (| ++ \rangle  + | -- \rangle )  \cr
  | \psi_{2}\rangle  &\equiv&  | \Phi^- \rangle = \frac {1} {\sqrt{2}}
  ( | +\frac {1} {2} +\frac {1} {2}\rangle  -
    |-\frac {1} {2} -\frac {1} {2} \rangle  ) \equiv
 \frac {1} {\sqrt{2}} (| ++ \rangle  - | -- \rangle )  \cr
  | \psi_{3}\rangle  &\equiv&  | \psi^+ \rangle = \frac {1} {\sqrt{2}}
  ( | +\frac {1} {2} -\frac {1} {2}\rangle  +
    |-\frac {1} {2} +\frac {1} {2} \rangle  )  \equiv
 \frac {1} {\sqrt{2}} (| +- \rangle  + | -+ \rangle )  \cr
  | \psi_{4}\rangle  &\equiv&  | \psi^- \rangle = \frac {1} {\sqrt{2}} 
  ( | +\frac {1} {2} -\frac {1} {2}\rangle  -
    |-\frac {1} {2} +\frac {1} {2} \rangle  )  \equiv
 \frac {1} {\sqrt{2}} (| +- \rangle  - | -+ \rangle ) ~.
\label{eq_1}
\end{eqnarray}

The elastic scattering of a proton beam off a proton target is described
by a transition operator $M$ \cite{book}, which can be expressed either
in terms of sixteen matrix elements
 $\langle m_1' m_2' \lvert M \rvert m_1 m_2 \rangle$
in the proton spin-projection basis $\rvert m_1 m_2 \rangle$, or in terms
of sixteen coefficients $C_{i'i}$  in the Bell basis (\ref{eq_1}):
\begin{eqnarray}
  M &=&  \sum_{m_1,m_2,m'_1,m'_2}  \langle m'_1 m'_2 | M | m_1 m_2 \rangle ~
  | m'_1 m'_2 \rangle \langle  m_1 m_2 | \cr
  &=& 
  \sum_{i,i'=1}^4  \langle \psi_{i'} | M | \psi_{i} \rangle ~
  | \psi_{i'} \rangle \langle \psi_i | \equiv 
   \sum_{i,i'=1}^4  C_{i'i} ~
  | \psi_{i'} \rangle \langle \psi_i |
   ~,
 \label{eq_6}
\end{eqnarray}
where the coefficients $C_{i'i}$ are determined  by  the matrix
elements $\langle m_1' m_2' \lvert M \rvert m_1 m_2 \rangle$ \cite{wit_unp_pd}.

The spin density matrix of the  outgoing protons is given in terms of the
transition operator $M$ and the spin state of the incoming protons, 
described by the density matrix
$\rho_{\text{in}}$, as \cite{ohlsen1972}:
\begin{eqnarray}
  \rho_f &=& M ~ \rho_{in} ~ M^{\dagger}
   ~.
 \label{eq_8}
\end{eqnarray}

In standard $pp$ scattering experiments, the initial spin states of the
proton beam and proton target are prepared independently, leading to
an initial spin density matrix given by the tensor product of the beam
and target spin density matrices. Each of these can be expressed in terms
of the corresponding polarization vectors of the beam, $P_i^{b}$, and
the target, $P_i^{t}$, $ i=1,2,3 ; (x,y,z)$ 
\cite{ohlsen1972}:
\begin{eqnarray}
  \rho_{\text{in}} &=& \frac{1}{4} (I^b + \sum_{i=1}^3 P_i^b \sigma_i)
  \otimes (I^t + \sum_{i=1}^3 P_i^t \sigma_i )    ~,
 \label{eq_8a}
\end{eqnarray}
where  $\sigma_i$ are the standard Pauli matrices.

For an unpolarized proton beam and target, the initial density matrix is
$\rho_{\text{in}} = \frac{1}{4} I^b \otimes I^t$  \cite{ohlsen1972}, and the
final spin density matrix, expressed in the Bell basis (\ref{eq_1}), is
characterized by sixteen coefficients $\bar C_{i'i}$  \cite{wit_unp_pd}:
\begin{eqnarray}
  \rho_f &\propto&  \sum_{i,i'=1}^4 ~
 (\sum_{i''=1}^4 C_{i'i''} C_{ii''}^*) ~
  | \psi_{i'} \rangle \langle \psi_i | \equiv 
   \sum_{i,i'=1}^4  \bar C_{i'i} ~
  | \psi_{i'} \rangle \langle \psi_i |
   ~.
 \label{eq_9}
\end{eqnarray}  

In Refs.~\cite{wit_unp_pd,shen_2025}, the properties of the $pp$ $M$ matrix
and the final-state spin density matrix $\rho_f$ in unpolarized $pp$
scattering were investigated. It was found that, at the center-of-mass
angle $\theta_{c.m.} = 90^\circ$, only three terms contribute to both
quantities. In addition, a strong dominance of a single term was observed
for both the $M$ matrix and $\rho_f$ at low energies (around $10$~MeV)
as well as at higher energies (within a narrow region around $150$~MeV).

Specifically, at lower energies, $E_{lab} \approx 10$~MeV, and for angles
around $\theta_{c.m.} \approx 90^\circ$, the transition matrix is well
approximated by
$M \approx | \psi^- \rangle \langle \psi^- |$,
while the corresponding final spin density matrix is given by
$\rho_f \approx | \psi^- \rangle \langle \psi^- |$.

At higher energies, $E_{lab} \approx 150$~MeV, the transition matrix
is instead approximated by
$M \approx | \psi^+ \rangle \langle \Phi^- |$,
and the corresponding final spin density matrix by
$\rho_f \approx | \psi^+ \rangle \langle \psi^+ |$.

The dominance of a single term in $\rho_f$ directly indicates that the
formation of strongly entangled Bell states in unpolarized $pp$ scattering
is possible \cite{wit_unp_pd}.
However, the energies at which such states can be produced are limited
either to the low-energy region (around $10$~MeV) or to a relatively narrow
region around $150$~MeV. Moreover, the angular region in which strongly
entangled Bell states are formed is centered around $\theta_{c.m.} = 90^\circ$.
At low energies, it spans a relatively broad range,
$\theta_{c.m.} \in (55^\circ,125^\circ)$, whereas around $150$~MeV it
becomes strongly restricted to $\theta_{c.m.} \in (85^\circ,95^\circ)$.
In both energy regions, a pure Bell state is produced only at
$\theta_{c.m.} = 90^\circ$. At other angles, the Bell states become
contaminated by contributions from other components, as evidenced
by the nonvanishing polarization of both entangled protons, equal to
a nonzero induced polarization.

The Bell states produced at low energies are of the $|\psi^{-}\rangle$
type, whereas in $pp$ elastic scattering at $150$~MeV the Bell-state
type changes to $|\psi^{+}\rangle$. In both cases, the energies of
the entangled protons are equal to one half of the incident proton energy
at $\theta_{\mathrm{lab}} = 45^\circ $, and become unequal as the scattering
angle deviates from this value.

This reduction in the energies of the outgoing entangled protons makes
$150$~MeV $pp$ scattering an unsuitable candidate for teleportation studies.
Although it produces an entangled proton pair, the energies of the
participating protons fall outside the region of energy where a single Bell term
dominates the $M$-matrix. As a result, the subsequent scattering process
required for teleportation can no longer be realized. For this reason, we
have restricted our investigation of the teleportation process to
the low-energy region.


\subsection{Numerical Simulation of a Quantum Teleportation Experiment}
\label{simul}

The most peculiar feature of proton scattering induced by the presence of
a strongly entangled Bell-like state in the initial configuration is
the possibility of teleportation of quantum mechanical spin
states \cite{teleport,wit_unp_pd}. This possibility is directly connected
with a novel feature that arises when working with entangled $pp$ Bell
states, namely the strong correlation between the proton spins, reflected
in a large value of the spin-correlation coefficient of $\pm 1$. This is
in contrast with standard scattering experiments, where the polarizations
of the participating particles in the initial state are prepared in a
completely independent manner.

Such large values of the spin correlation imply that the results of
spin-projection measurements performed on the two protons are perfectly
correlated: the result of a spin measurement on one proton uniquely
determines the outcome of the corresponding measurement on the other
proton. This leads to interesting consequences when one of the entangled
protons is scattered off a polarized hydrogen target at energies and
scattering angles for which a single Bell term dominates the $pp$
scattering matrix.

Under such conditions, the final spin state of the second proton in the Bell
pair becomes identical to the spin state of the hydrogen target, while
the scattered proton pair emerges in a Bell state of the same type as the
initial one. This process may be interpreted as the teleportation of
the quantum-mechanical spin state from the hydrogen target to the initially
entangled proton.

However, this should not be interpreted as a simultaneous teleportation of
the initial Bell state to the scattered proton pair. The formation of the
two entangled pairs is completely independent, and the Bell-state structure
of the scattered pair arises solely from the dominance of a single
Bell-state contribution in the corresponding transition matrix $M$.
Furthermore, the admixture of other Bell-state components, as evidenced by
the nonvanishing polarizations of both entangled protons, originates from
the dynamics of the two scattering processes at different incident
proton energies.

For the convenience of the reader, we again show in Fig.~\ref{fig1}
the kinematics of scattering of the three protons and recall some formulas
from Ref.~\cite{wit_unp_pd}. In the ideal case where the $pp$ pair $23$
is in a pure Bell state $| \psi^- \rangle_{23}$ and the polarized hydrogen
target $1$ has polarization $P_y^1$, the initial state is described by
the spin density matrix~\cite{ohlsen1972}:
\begin{eqnarray}
  \rho_{123}^{\text{in}} =    \rho_{\text{23}} \otimes \rho_{\text{1}} =
 | \psi^- \rangle_{23}   ~_{23}\langle \psi^- |  \otimes 
  \frac{1}{2} ( I^1 + P_y^1 \sigma_y^1 )     
   ~.
 \label{eq_10}
\end{eqnarray}    

The final density matrix of the system after proton 2 scatters off proton 1 is:
\begin{eqnarray}
  \rho_f &=& M_{12} \otimes I^3 ~  \rho_{123}^{\text{in}} ~
  (M_{12} \otimes I^3 )^{\dagger}
   ~.
 \label{eq_11}
\end{eqnarray}

Taking the form $M_{12} = | \psi^- \rangle_{12} ~_{12}\langle \psi^- |$,
which is approximately valid at $E_{lab} \approx 10$~MeV, a direct
calculation leads to~\cite{wit_unp_pd}:
\begin{eqnarray}
  \rho_f &=&   | \psi^- \rangle_{1'2'}   ~ _{1'2'}\langle \psi^- |
  \otimes \frac {1} {2} ( I^3 + P_y^1 \sigma_y^3 )
  ~,
 \label{eq_12}
\end{eqnarray}
demonstrating that the spin state of the target proton $1$ is indeed
teleported to proton $3$, while the scattered proton pair $1'2'$ emerges
in a Bell state of the same type as the initial pair $23$.

However, since the above formulas were derived under the assumption that
only a single Bell term contributes to the transition matrix $M$, this
conclusion is valid only in the case where
$\theta_{lab}^2 = \theta_{lab}^3 = \theta_{lab}^{2'} = 45^\circ$. In this
configuration, the induced polarization in $pp$ scattering vanishes
due to the identity of the protons, and both proton pairs form pure Bell states.

For scattering angles different from $45^\circ$, nonvanishing contributions
from induced polarizations and polarization transfers appear, leading to
nonzero polarization of the outgoing protons. As a consequence, the
resulting proton pairs are no longer in pure Bell states.

In order to investigate how deviations from the dominance of a single Bell
term in the transition matrix, as well as contamination of the Bell state
by additional contributions, influence the teleportation process, we
performed numerical simulations of a three-proton system. This was done
by determining all relevant final spin observables through calculation
of the final density matrix of the system.

To this end, we solved the Lippmann–Schwinger equation for
$pp$ scattering~\cite{book} using the high-precision AV18
NN potential~\cite{av18}, with the Coulomb interaction between
the two protons included explicitly. 
This allowed us to determine the required transition
matrix $M_{23}(\theta_{2})$ at the energy of the incoming unpolarized
proton $E_{lab}$, as well as all transition matrices
$M_{12}(\theta_{2},\theta_{2'})$ required for subsequent scatterings of
proton $2$ on the target proton $1$ at different angles $\theta_2$ and,
consequently, at different energies of proton $2$.

We performed this analysis at four incident energies, $E_{lab} = 5, 10, 15$,
and $20$~MeV, and present in Figs.~\ref{fig2}–\ref{fig7} the predictions
at $E_{lab} = 10$~MeV for the final polarizations
$\langle \sigma_y^{i'} \rangle$ and spin correlations
$\langle \sigma_y^{i'} \sigma_y^{j'} \rangle$ ($i',j' = 1,2,3$) as functions
of the center-of-mass angles $\theta^{2}_{c.m.}$ and $\theta^{2'}_{c.m.}$.
The results at the other energies (not shown) are qualitatively very similar.

In each figure we present two panels: one for an unpolarized hydrogen
target $1$ (upper panel a)), and one for a polarized target
with $P_y^1 = 0.1$ (lower panel b)).

Let us first examine spin correlations, whose values close to $\pm 1$
provide strong evidence of entanglement. It was shown in
Ref.~\cite{wit_unp_pd} that strongly entangled proton pairs are formed,
and that a dominance of a single Bell term in the transition matrix $M$ and
in the spin density matrix $\rho$ occurs in the region of c.m. angles
$\theta_{c.m.} \in (55^\circ, 125^\circ)$ (see Fig. 28a and Table I
in Ref.~\cite{wit_unp_pd}).

As can be seen in Fig.~\ref{fig2}, in this angular region of
$\theta^{2}_{c.m.}$ and $\theta^{2'}_{c.m.}$, the spin correlation of
the pair $1'2'$ approaches $-1$. This indicates that the dominance of a
single Bell-state contribution in $M_{12}$ leads to the formation of a
strongly entangled pair $1'2'$, just as the dominance of such a contribution
in $M_{23}$ led to the formation of the strongly entangled initial pair $23$
in the first scattering of an unpolarized proton from an unpolarized target
(see Fig.~28a in Ref.~\cite{wit_unp_pd}).

At the same time, this scattering results in a strong reduction of the
spin correlation within the pair $2'3'$, as shown in Fig.~\ref{fig4}.
The spin correlation of the pair $1'3'$ (Fig.~\ref{fig3}) is, as expected,
small and very similar to that of the pair $2'3'$.

A comparison of the spin correlations displayed in
Figs.~\ref{fig2}a)--\ref{fig4}a) and Figs.~\ref{fig2}b)--\ref{fig4}b)
reveals that the polarization of target $1$ has a negligible effect on
the spin-correlation values.

Teleportation is expected to affect the final polarizations,
in particular the polarization of proton $1$ should be transferred to
proton $3$. Consequently, for the case of an unpolarized hydrogen
target $1$, one would conclude that the effects of teleportation  cannot
be  identified by examining  the polarization of the
outgoing proton $3'$, which should vanish in case of pure Bell states.

Since the pair $1'2'$ emerges in a strongly correlated Bell-like state,
the polarizations of $1'$ and $2'$ in the relevant region of c.m. angles
should approach zero, as shown in Figs.~\ref{fig5}a) and \ref{fig6}a).
Suprisingly, polarization of proton $3'$ does not vanish for unpolarized
target $1$ but remains small, as seen in Fig.~\ref{fig7}a).

For all three protons, the polarizations $\langle \sigma_y^{i'} \rangle$,
although small in magnitude, vary smoothly as functions of the angles
$\theta^{2}_{\mathrm{c.m.}} $ and $ \theta^{2'}_{\mathrm{c.m.}} $.
Their behavior closely resembles that of the induced polarization in $pp$
scattering. For protons $1'$ and $2'$, the polarization depends primarily
on $ \theta^{2'}_{\mathrm{c.m.}} $, whereas for proton $3'$ it depends
mainly on $ \theta^{2}_{\mathrm{c.m.}} $, being practically independent
of $ \theta^{2'}_{\mathrm{c.m.}} $. In this respect, it closely resembles
the induced polarization of proton $3$ in the first scattering.

As can be clearly seen in Figs.~\ref{fig5}b) and \ref{fig6}b),
a nonvanishing polarization of target $1$ affects the polarizations
of protons $1'$ and $2'$ more strongly than it affects the spin correlations.
Nevertheless, these polarizations remain small throughout the angular
region where entanglement occurs.

For proton $3'$, in contrast, a clear signature of teleportation is observed.
As shown in Fig.~\ref{fig7}b), the polarization of target $1$ is completely
transferred to proton $3'$ over the entire angular region of entanglement.

In order to better understand the behaviour of the presented observables and
mechanism of teleportation, we
analyzed the interaction of three protons using the standard language of
spin formalism in nuclear physics. In Appendix~\ref{a1}, expressions for
the final polarizations and spin correlations in terms of induced
polarizations, spin correlations, and polarization (spin-correlation)
transfer coefficients are given
\cite{ohlsen1972,wit_spin_doubl,wit_spin_np_entangl}, under the assumption
that the  only polarized particle in the initial configuration is
the hydrogen target $1$, with a nonvanishing polarization component
$P_y^1$.

Assuming further the dominance of a single Bell term in the contributing
transition matrices $M$, the  induced polarizations,
induced spin correlations, spin-correlation
transfer coefficients, as well as final polarizations and spin-correlations
 for three participating protons and three possible pairs of them 
are evaluated, providing the final observables.
Such domination in both transition matrices $M_{12}$ and $M_{23}$  
occurs at energies $E_{lab} \approx 10$~MeV to a very good approximation
in the angular region of entanglement 
$\theta_{c.m.}^2$ and  $\theta_{c.m.}^{2'} \in (55^\circ,125^\circ)$. 

Let us first discuss, in light of the results obtained in Appendix~\ref{a1},
the final polarizations of protons $1'$ and $2'$ in this  region of angles.
The induced contribution, $P_y^{1'(2')~ \mathrm{ind}}$, to the final polarization
of protons $1'$ and $2'$ in Eq.~(\ref{eq_ape4}) is given by the induced
polarization in the scattering of proton $2$ on proton $1$
(see Eqs.~(\ref{eq_ape7.4}), (\ref{eq_ape7.7}), and
Refs.~\cite{ohlsen1972,wit_spin_np_entangl}).
Likewise, the contribution determined by the polarization-transfer
coefficients $K_y^{y'}(1 \to 1'(2'))$ is, for both protons $1'$ and $2'$,
given by the corresponding polarization-transfer coefficient in $pp$
scattering at the center-of-mass scattering angle
$\theta_{c.m.}^{2'}$ (see Eqs.~(\ref{eq_ape7.5}) and (\ref{eq_ape7.7})).
In both cases, these
observables correspond to $pp$ scattering at the energy of the
incoming proton $2$  at the scattering angle $\theta_{c.m.}^{2}$.

For an unpolarized hydrogen target, $P_y^1 = 0$, the final polarizations
$\langle \sigma_y^{1'} \rangle$ and $\langle \sigma_y^{2'} \rangle$
coincide with the induced polarization in the scattering of proton $2$
from proton $1$. This accounts for the behaviour observed in Figs.~\ref{fig5}a)
and \ref{fig6}a), including the dependence on $\theta_{c.m.}^2$, which reflects
the change in the incident energy of proton $2$, and on $\theta_{c.m.}^{2'}$,
which governs the magnitude of the induced polarization.

For a polarized hydrogen target, the final polarizations
$\langle \sigma_y^{1'} \rangle$ and $\langle \sigma_y^{2'} \rangle$ receive an 
additional contribution from the polarization transfer
coefficients $K_y^{y'}(1 \to 1'(2'))$:
$$  \langle \sigma_y^{1'(2')} \rangle =   
   P_y^{1'(2')~ind} +  P_y^1 K_y^{y'}(1 \to 1'(2')) ~.  $$
   This follows from the fact that, in the angular region of strong
   entanglement, the  trace appearing 
   in Eq.~(\ref{eq_ape7.6}) vanishes,  
$$Tr (M_{12} M_{23}M_{23}^{\dagger}  \sigma_y^1 M_{12}^{\dagger} ) = 0 ~, $$ 
and therefore $$Tr(\rho_f) = Tr( \rho_f^0).$$

For proton $3'$, the induced contribution to its final polarization
in Eq.~(\ref{eq_ape4}), $P_y^{3'~ \mathrm{ind}}$, is given by
Eqs.~(\ref{eq_ape7.2}) and (\ref{eq_ape7.7}). In the angular region
where entanglement occurs, this quantity coincides with the induced
polarization in the first $pp$ scattering.

Furthermore, in this angular region the polarization-transfer coefficient
reaches its maximal value, $K_y^{y'}(1 \to 3') = 1$
(see Eqs.~(\ref{eq_ape7.3}) and (\ref{eq_ape7.1})). As a result,
the final polarization of proton $3'$ takes the form
$$
\langle \sigma_y^{3'} \rangle = P_y^{3'~\mathrm{ind}} + P_y^1.
$$

 The condition $K_y^{y'}(1 \to 3') = 1$ is a clear signature of the
 teleportation process and can be directly linked to the dominance of
 a single Bell-state contribution in the transition matrix $M_{12}$ describing
 the scattering of proton $2$ from proton $1$ (see Eq.~(\ref{eq_ape7.3})).

 This implies that, for an unpolarized target proton $1$, the final
 polarization of proton $3'$ coincides with the induced polarization
 generated in the first scattering. Accordingly, the results shown
 in Fig.~\ref{fig7}a) are independent of $\theta_{c.m.}^{2'}$ and depend
 only on $\theta_{c.m.}^{2}$. Therefore, measuring this polarization does
 not provide information about the teleportation process and cannot
 be regarded as its signature.

The final spin correlations between different proton pairs are determined
by the induced correlations,
$ \langle \sigma_y^{i'} \sigma_y^{j'} \rangle^{\mathrm{ind}} $, and the single-spin
correlation-transfer coefficients $ K_{0y}^{y'y'}(1 \to i'j') $ appearing
in Eq.~(\ref{eq_ape8}). For an unpolarized hydrogen target, $ P_y^1 = 0 $,
the spin correlations are entirely determined by the induced contribution.

When a single Bell-state term dominates the matrix $ M_{23} $, the induced
contribution coincides with that obtained in the scattering of an
unpolarized proton $2$ from an unpolarized hydrogen target $1$
(see Eq.~(\ref{eq_ape12})). In the same limit, the coefficient
$ K_{0y}^{y'y'}(1 \to i'j') $ is also identical to its counterpart in
that scattering process.

When the dominance of a single Bell-state term also occurs in $M_{12}$,
the induced spin correlation
$\langle \sigma_y^{1'} \sigma_y^{2'} \rangle^{\mathrm{ind}}$ becomes 
$\langle \sigma_y^{1'} \sigma_y^{2'} \rangle^{\mathrm{ind}} = -1$, and the
proton pair $1'2'$ is in a strongly entangled Bell-like state
in all angular region of entangelement as seen in Fig.~\ref{fig2}a).
  It is again evident that the dominance of a single Bell-state contribution
  in the transition matrix $M_{12}$ is responsible for the formation of
  the strongly entangled pair $1'2'$.

Since, in this angular region, the induced spin correlations
$\langle \sigma_y^{1'} \sigma_y^{3'} \rangle^{\mathrm{ind}}$ and
$\langle \sigma_y^{2'} \sigma_y^{3'} \rangle^{\mathrm{ind}}$ vanish, the
behaviour shown in Figs.~\ref{fig3}a) and \ref{fig4}a) can be attributed to
the dominance of a single Bell-state component in the transition matrices
$M_{12}$ and $M_{23}$. The negligible influence of the target proton $1$
polarization on the spin-correlation observables
(see Figs.~\ref{fig2}b)--\ref{fig4}b)), arising from vanishing
single-spin correlation-transfer coefficients $K_{0y}^{y'y'}(1 \to i'j')$,
is likewise consistent with this mechanism.

  It should be emphasized that the entangled states of the pairs $23$ and $1'2'$
  are not identical. The difference arises from entanglement-degrading
  contributions associated with the nonzero polarizations of
  their constituent protons.

  For the pair $23$, these contributions are determined by the induced
  polarizations generated by the transition matrix $M_{23}$, whereas
  for the pair $1'2'$ they are governed by $M_{12}$. Because the energy of
  the incident proton $2$ differs from that of the unpolarized proton in
  the first scattering, the corresponding induced polarizations—and hence
  the deviations from a pure Bell state—are also different.

  Polarizing the target proton $1$ modifies the final polarizations of all
  three protons (see Figs.~\ref{fig5}b–\ref{fig7}b). For protons $1'$
  and $2'$ (Figs.~\ref{fig5}b and \ref{fig6}b), the changes are relatively
  small, and their polarizations remain close to zero in the angular region
  where strong entanglement occurs. This can be attributed to the small
  values of the polarization-transfer coefficients $K_y^{y'}(1 \to 1')$
  and $K_y^{y'}(1 \to 2')$ in low-energy $pp$ scattering.

  In contrast, the polarization of proton $3'$ changes dramatically, exhibiting
  a clear signature of the teleportation of the spin state of proton $1$
  to proton $3'$ (Fig.~\ref{fig7}b).

Throughout the angular region defined by $\theta_{c.m.}^2$ and
$\theta_{c.m.}^{2'}$, where strong entanglement is present, the polarization
of proton $3'$ is equal to the target polarization $P_y^1$, with only
a small correction arising from the induced term $P_y^{3'~\mathrm{ind}}$.

We further investigated the dependence of the teleportation effect on
the polarization of proton target  $1$. We found that varying the target
polarization from small values of $P_y^1$ up to its maximum value,
$P_y^1 = 1$, consistently produces a clear teleportation signal within
the angular region of strong entanglement. Remarkably, the effect remains
observable even for polarization values as small as $P_y^1 = 1\%$.

This behavior is illustrated in Fig.~\ref{fig8} for two low values of
the target polarization. The case $P_y^1 = 0.034$, shown in Fig.~\ref{fig8}b,
corresponds to a realistic polarized hydrogen target, as described in
Refs.~\cite{watanabe,tateishi}.

It is clear that the most conclusive evidence of teleportation would be
provided by a measurement of the polarization of proton $3'$ using
a polarized target proton $1$. Since the realization of such an experiment
does not currently appear feasible, one must restrict the discussion
to the case of an unpolarized hydrogen target $1$.

However, in this case teleportation does not take place at all, and what
remains are residual spin correlations in the nucleon pair $1'2'$ arising
from strong entanglement.

The most direct confirmation of these correlations in the three-proton
system would be provided by a measurement of the final
spin-correlation coefficient
$\langle \sigma_y^{1'} \sigma_y^{2'} \rangle^{\mathrm{ind}}$.
A simpler alternative would be to measure the final polarization of
proton $2'$, comparing the cases with the unpolarized hydrogen
target $1$ present and removed.

In the latter case, when the target is removed, the final polarization of
proton $2'$ is determined solely by the induced polarization generated in
the first scattering. When the target is present, the final polarization
is determined by the induced polarization acquired by proton $2'$
in the second scattering from the hydrogen target $1$.

An observable difference between these two polarizations would constitute
indirect evidence for the formation of the strongly correlated pair $1'2'$.

Figure~\ref{fig9} shows, for incident proton energies of
$E_{\mathrm{lab}} = 10$, $15$, and $20$~MeV, the final polarization of proton $2$,
$\langle \sigma_y^{2} \rangle$, resulting from the first scattering with
the unpolarized target $1$ removed (red solid line), plotted as a function
of the center-of-mass scattering angle $\theta_{\mathrm{c.m.}}^{2}$.

Also shown is the final polarization of proton $2'$,
$\langle \sigma_y^{2'} \rangle$, after the second scattering of proton $2$
from target $1$, plotted as a function of $\theta_{\mathrm{c.m.}}^{2'}$.
Results are presented for two laboratory scattering angles of proton $2$:
$\theta_{\mathrm{lab}}^{2}=45^\circ$ (blue dashed line) and
$\theta_{\mathrm{lab}}^{2}=65^\circ$ (black dotted line).

The approximately threefold difference between these polarization values,
together with the pronounced dependence of $\langle \sigma_y^{2'} \rangle$
on $\theta_{\mathrm{c.m.}}^{2'}$, suggests that the effect may be experimentally
observable. Nevertheless, the small magnitudes of the polarizations,
particularly that of proton $2'$, impose stringent requirements on
the precision of their measurement.

Quantum teleportation is an interesting process in which the interaction of
one member of an entangled pair of protons with an external proton leads to
the formation of an entangled pair consisting of the interacting protons,
while leaving the second proton of the original pair in the state of the
external proton. In the next section, we will see that this is a special
case of a more general process in which the interaction of two protons
belonging to different entangled pairs leads not only to the formation
of an entangled pair of the interacting protons, but also simultaneously
produces an entangled pair formed by their noninteracting counterparts.

\subsection{Transfer of Entanglement Between Proton Pairs}
\label{transfer}

In the following, we discuss features of proton scattering, other than
teleportation, that arise from the presence of a strongly entangled
Bell-like state in the initial configuration,
assuming that the dominance of a single Bell component is retained
in both the transition matrix $M$ and the spin density matrix $\rho$ at
the energy of the outgoing protons. 
This restriction excludes from consideration entangled $pp$ pairs formed
at approximately $150$~MeV, since at the energy of the scattered protons,
which is about $75$~MeV for a scattering angle of $\theta_{lab}=45^\circ$,
neither the $M$ matrix nor the spin density matrix $\rho$ is dominated
by a single term in the Bell basis.

For an initial proton energy of about
$10$~MeV, the transition matrix is well approximated by
$M \approx |\psi^-\rangle\langle\psi^-|$, \cite{wit_unp_pd}. 
Here, the reduction of the outgoing  protons energies at a scattering angle of
$\theta_{lab}=45^\circ$ by a factor of two
relative to the incoming proton energy is actually advantageous.
Specifically, the dominance of a single term in the $M$ matrix, as well as
the quality of the generated entangled states, is enhanced by lowering the
energies of the outgoing protons
(see Figs.~1 and 2 in Ref.~\cite{wit_unp_pd}).
Consequently, successive scatterings of protons from entangled pairs
on an unpolarized hydrogen target will generate an increasingly growing 
network of entangled proton pairs, as depicted in Fig.~\ref{fig1a}.

Let us consider an entangled proton pair, labeled $3$ and $4$ in
the network, prepared in the Bell state $ |\psi^- \rangle $ and produced
in proton $1$  off proton $2$  scattering, with each
outgoing proton emerging at a
laboratory angle of $\theta_{\mathrm{lab}} = 45^\circ$ (see Fig.~\ref{fig1a}).
Since the energies of the outgoing protons are equal to half the energy
of the incoming proton, subsequent scatterings in which the outgoing
protons again emerge at the same angle are characterized by $M$ and $\rho$
operators that are increasingly dominated by a single Bell-state contribution.
Consequently, one may approximate $M \approx |\psi^- \rangle \langle \psi^- $, 
as discussed in Ref.~\cite{wit_unp_pd}.

In the subsequent step, the entangled protons $3$ and $4$ are scattered
off unpolarized hydrogen targets $5$ and $6$, respectively. The initial
spin density matrix of the four-proton system can be written as:
\begin{eqnarray}
  \rho_{\text{in}} =    \rho_{\text{34}} \otimes \rho_{\text{5}}
  \otimes \rho_{\text{6}} =
 | \psi^- \rangle_{34}   ~_{34}\langle \psi^- |  \otimes 
  \frac{1}{2} I^5  \otimes   \frac{1}{2} I^6 
   ~,
 \label{eq_ap1}
\end{eqnarray}
where the initial state of the entangled protons $3$ and $4$ is the Bell state
$ |\psi^- \rangle_{34} $.

The final density matrix of the system following the scattering of proton $3$
off proton $5$ and proton $4$ off proton $6$ is:
\begin{eqnarray}
  \rho_f &=& M_{35} \otimes M_{46} ~  \rho_{\text{in}} ~
  (M_{35} \otimes M_{46})^{\dagger}
   ~.
 \label{eq_ap2}
\end{eqnarray}
Taking $M_{35}=|\psi^-\rangle_{35} ~_{35}\langle\psi^-|$ and 
 $M_{46}=|\psi^-\rangle_{46} ~_{46}\langle\psi^-|$, 
which are valid approximations at $E_{\mathrm{lab}} \approx 10$ MeV,
a straightforward calculation yields:
\begin{eqnarray}
  \rho_f &=&  \frac {1} {4} | \psi^- \rangle_{79}   ~_{79}\langle \psi^- |
  \otimes | \psi^- \rangle_{810}   ~_{810}\langle \psi^- |
   ~.
 \label{eq_ap3}
\end{eqnarray}

Thus, following the scatterings, the system evolves into a new state
in which protons $7$ and $9$ are prepared in the Bell state
$ |\psi^- \rangle_{79} $, while protons $8$ and $10$ are prepared in the Bell
state $ |\psi^- \rangle_{810} $.

To gain further insight into the mechanism responsible for the double
production of entangled states described above, it is useful to consider
the process analyzed in Ref.~\cite{wit_unp_pd}. In that process, proton $3$
from the entangled $pp$ pair $34$ is scattered from an unpolarized hydrogen
target $5$. As a result, the outgoing proton pair $79$ is produced in
the same Bell state as the initial pair $34$, while proton $4$
emerges unpolarized.

Subsequently, the now unpolarized proton $4$ undergoes scattering from an
unpolarized hydrogen target $6$. Since this process is identical to the
scattering of proton $1$ from proton $2$, it results in the formation of
two entangled pairs, $79$ and $810$, both in Bell states of the same type
as the initial pair $34$.

The same final configuration can, of course, be generated by considering
the analogous sequence of reactions initiated by proton $4$ rather
than proton $3$.

Another interesting feature of the mechanism is its ability to transfer
entanglement between members of different entangled $pp$ pairs.
Consider the entangled pairs $1314$ and $1516$ shown in Fig.~\ref{fig1a}.
When proton $14$ is scattered off proton $15$ under the kinematical
conditions displayed in Fig.~\ref{fig2a}, the outgoing pair $14'15'$
becomes entangled, while entanglement is simultaneously established
between protons $13$ and $16$.

The initial spin density matrix of the four-proton system $(13,14,15,16)$
is given by:
\begin{eqnarray}
  \rho_{\text{in}} =    \rho_{\text{1314}} \otimes \rho_{\text{1516}} =
 | \psi^- \rangle_{1314}   ~_{1314}\langle \psi^- |  \otimes 
 | \psi^- \rangle_{1516}   ~_{1516}\langle \psi^- |
   ~,
 \label{eq_ap1a}
\end{eqnarray}
and the $M$ operator,
$M=| \psi^- \rangle_{1415}   ~_{1415}\langle \psi^- |\otimes I^{13} |
\otimes I^{16}$, 
yields the final density matrix of the system:
\begin{eqnarray}
  \rho_f &=&  \frac {1} {4} | \psi^- \rangle_{14'15'}   ~_{14'15'}\langle \psi^- |
  \otimes | \psi^- \rangle_{1316}   ~_{1316}\langle \psi^- |
   ~.
 \label{eq_ap3a}
\end{eqnarray}

The above result holds  when the scattering of protons $14$ and $15$,
viewed in their respective laboratory frames (i.e., with either proton $14$
or $15$ at rest), produces two outgoing protons emerging in these frames
at a laboratory angle of $\theta_{\mathrm{lab}} = 45^\circ$.
These kinematical conditions satisfy the requirement that a single Bell term
dominates in the $M$ and $\rho$ matrices.

That this is indeed the case is illustrated in Fig.~\ref{fig2a} for the
laboratory frame of proton $14$, obtained via the corresponding
Galilean transformation. After the scattering in the coordinate system
shown in Fig.~\ref{fig1a}, proton $14'$ remains at rest, while the second
proton $15'$ moves upward with energy twice that of each incoming proton.

It is evident that such a transfer of entanglement to protons belonging to
multiple distinct entangled pairs can be achieved by performing
appropriate scatterings between constituent protons from different pairs.
However, this appears to be of purely academic interest, since even
in the case of the two pairs considered above, its experimental
realization seems unlikely.

While scattering between protons originating from different entangled
pairs leads to entanglement transfer, scattering between the constituents
of a given pair leaves the entanglement intact.

\section{Summary and Conclusions}
\label{sumary}

We investigated the low-energy teleportation of a quantum-mechanical state
in a three-proton system with the aim of identifying
a simple and experimentally feasible signature of this process.

To gain a better understanding of the underlying reaction mechanisms
and the role of spin degrees of freedom, we formulated the problem
within the standard spin formalism commonly used in nuclear physics.
In addition, to remain as close as possible to realistic physical conditions,
we performed numerical simulations of the teleportation process using a
realistic proton–proton interaction and calculated the corresponding
spin observables.

The analysis of the reactions leading to teleportation in a three-proton
system provides unambiguous evidence that, as expected, the dominance of
a single Bell component in the transition matrices $M$ of the two
contributing $pp$ scattering processes is responsible for the occurrence
of teleportation.

The transition matrix $M_{23}$ of the first $pp$ scattering generates
strongly entangled Bell-like states over a broad range of scattering
angles $\theta_{c.m.}^{2}$ and $\theta_{c.m.}^{2'}$. The process of teleportation
itself is, however, directly driven by the transition matrix $M_{12}$
associated with the second scattering, in which proton $2$ interacts
with hydrogen target $1$.

The dominance of a single Bell component in this transition matrix leads
directly to the teleportation of the polarization of target $1$ to
proton $3'$ and to the formation of a strongly correlated $1'2'$ pair.
In the standard spin formalism, this corresponds to a polarization-transfer
coefficient $$K_y^{y'}(1 \rightarrow 3')=1,$$ 
and an induced spin correlation
$$\langle \sigma_y^{1'} \sigma_y^{2'} \rangle^{\mathrm{ind}}=-1.$$

Numerical simulations of the teleportation process have shown that, even for
small values of the polarization of proton $1$, this polarization is
faithfully teleported to proton $3'$ throughout the entire region of
scattering angles $\theta_{c.m.}^{2}$ and $\theta_{c.m.}^{2'}$ characterized
by strong entanglement. Measurement of the final polarization of proton $3'$
would provide clear evidence that quantum teleportation has occurred.

When the polarization of proton $1$ vanishes, the teleportation process
ceases to exist and no longer manifests itself in the final-state
polarizations. In the region of strong entanglement, the polarizations
of all outgoing protons are determined solely by the induced polarizations
associated with the corresponding $pp$ scattering processes: the scattering
of proton $2$ from proton $1$ in the case of protons $1'$ and $2'$, and
the scattering of an unpolarized incident proton from an unpolarized
proton target in the case of proton $3'$.

Consequently, for an unpolarized target proton $1$, no  experimental
signature of teleportation remains. The only unambiguous evidence of
the underlying quantum correlations is the formation of a strongly
entangled and spin-correlated proton pair $1'2'$. Establishing the
presence of this entanglement therefore requires direct measurements
of the relevant spin-correlation observables.

A simpler experimental approach would be to compare the final polarization
of proton $2'$ in the presence and absence of the unpolarized hydrogen
target proton $1$. Although such a measurement would not provide direct
evidence for the formation of the strongly entangled pair $1'2'$, it could
serve as an indirect signature of its presence. Owing to the small expected
polarization values, however, achieving the required experimental
accuracy may be challenging.

The process involving an unpolarized target proton $1$ provides valuable
insight into the mechanism by which a second strongly entangled Bell pair
is generated through the scattering of proton $3'$ from an unpolarized
target. It also offers a natural framework for understanding entanglement
transfer between protons, whereby quantum correlations are redistributed
among the particles through their mutual interactions.

\appendix

\section{Final Polarizations and Spin Correlations}
\label{a1}

\begin{figure}
  \includegraphics[bb=0 123 405 570, scale=1.0]{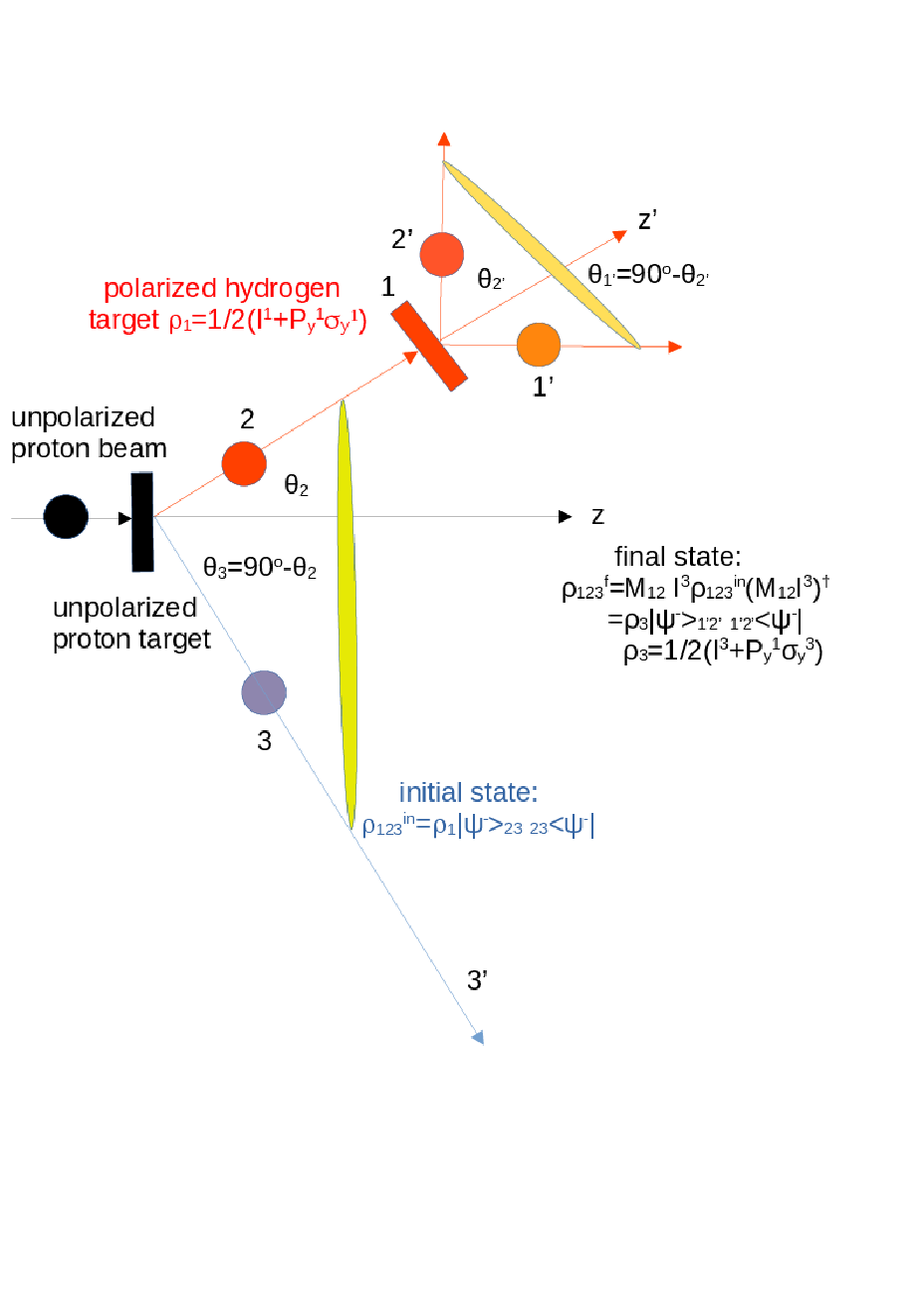}
  \caption{Production of an entangled proton pair $23$ and teleportation
    of the spin state of proton $1$ to proton $3'$ via the scattering
    of proton $2$ from a polarized hydrogen target $1$.
  }
\label{fig1}
\end{figure}

\begin{figure}
\begin{center}
\begin{tabular}{c}
\resizebox{120mm}{!}{\includegraphics[angle=270]{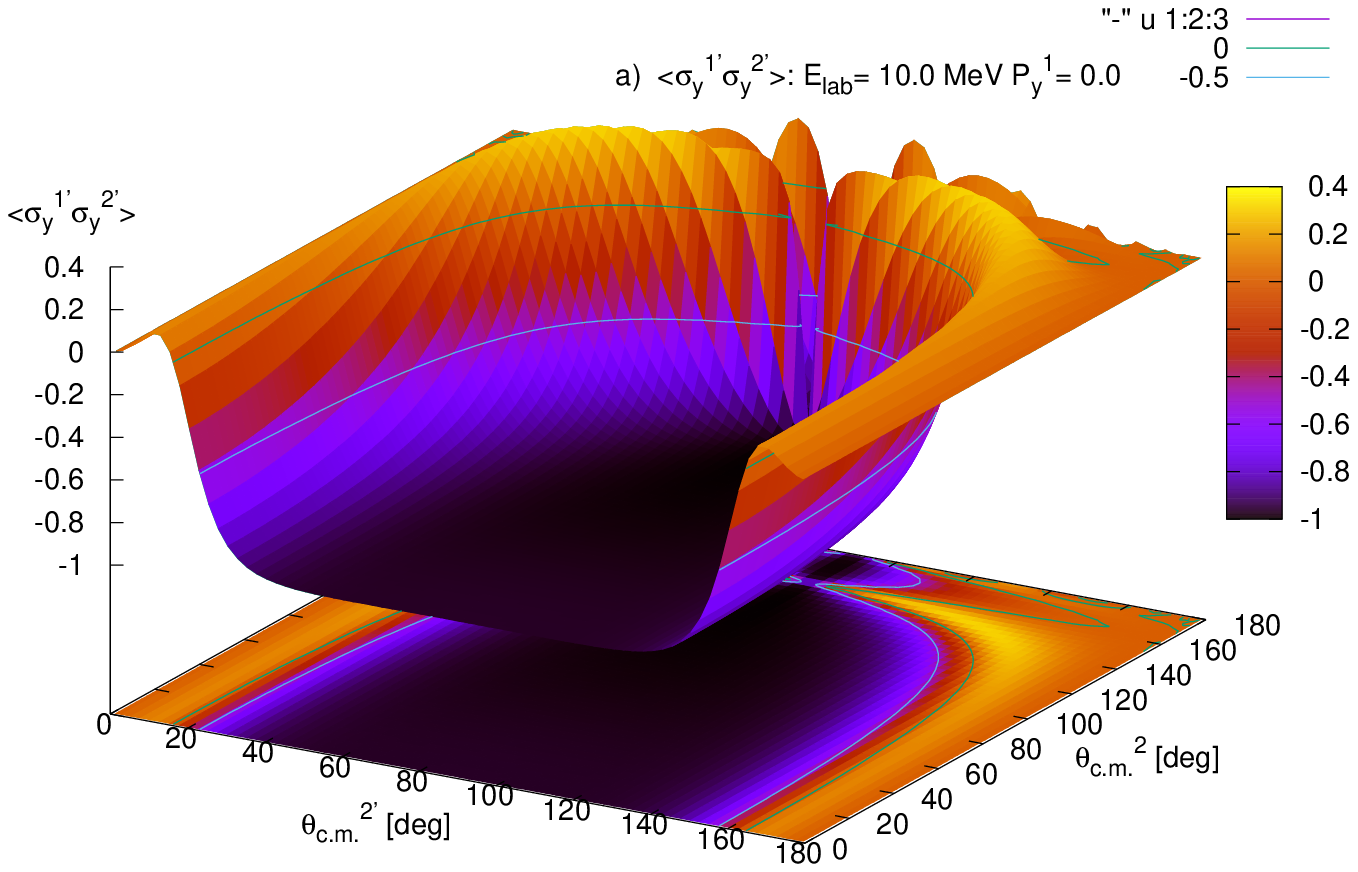}} \\
\resizebox{120mm}{!}{\includegraphics[angle=270]{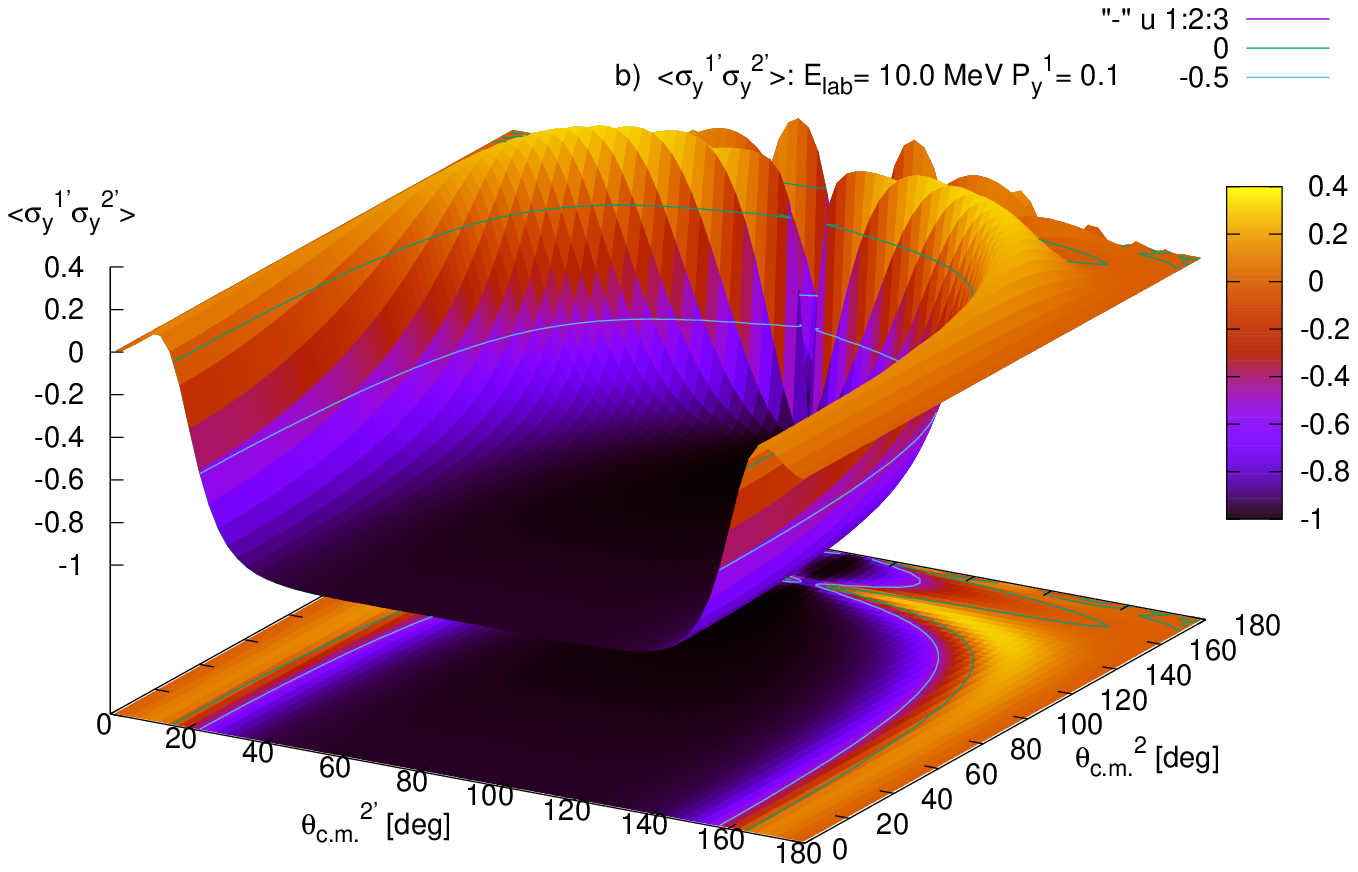}} \\
\end{tabular}%
\caption{
  (color online)
  Spin correlation of protons $1'2'$,
  $\langle \sigma_y^{1'} \sigma_y^{2'} \rangle$. The incident unpolarized
  proton, whose scattering from an unpolarized hydrogen target produces
  the entangled pair $23$, has laboratory energy $E_{\mathrm{lab}} = 10$~MeV.
  $\theta_{\mathrm{c.m.}}^{2}$ denotes the c.m. scattering angle of proton $2$.
  $\theta_{\mathrm{c.m.}}^{2'}$ is the c.m. angle of proton $2'$ in the
  subsequent scattering of proton $2$ from an unpolarized (a))
  or polarized (b)) hydrogen target $1$. Calculations use the AV18
  potential with $j_{\mathrm{max}} = 5$.
}  
\label{fig2}
\end{center}
\end{figure}

\begin{figure}
\begin{center}
\begin{tabular}{c}
\resizebox{135mm}{!}{\includegraphics[angle=270]{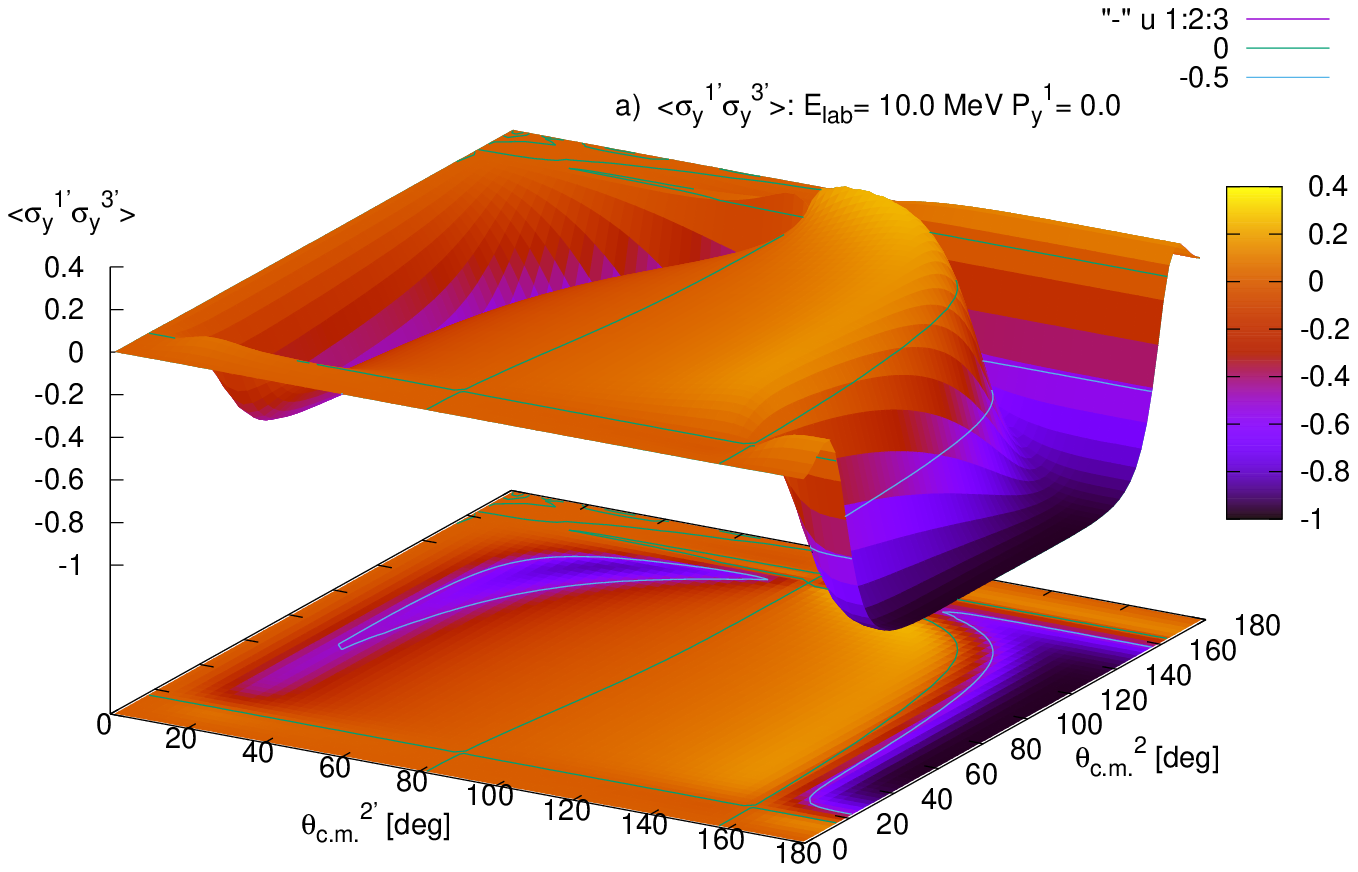}} \\
\resizebox{135mm}{!}{\includegraphics[angle=270]{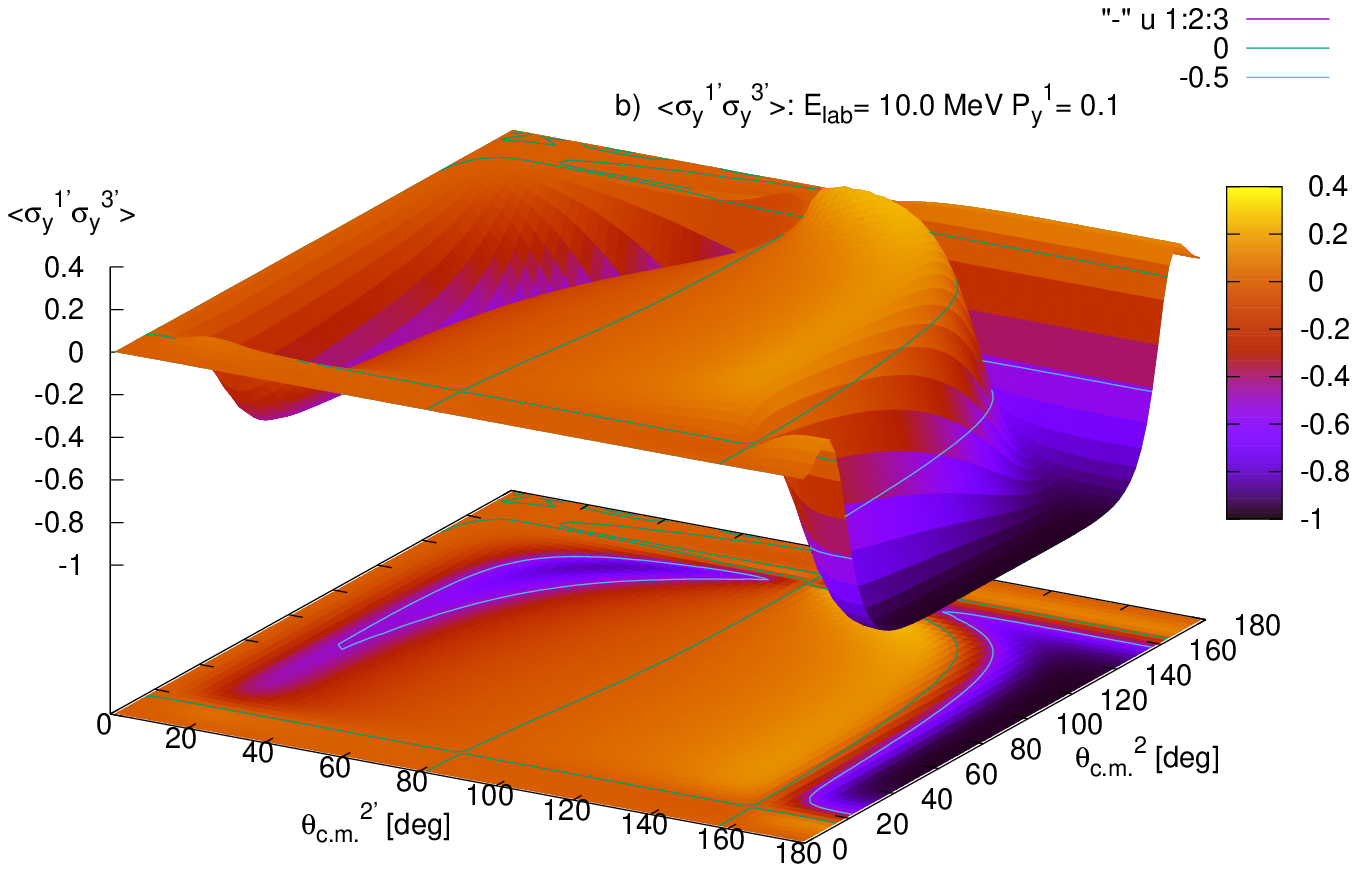}} \\
\end{tabular}%
\caption{
  (color online)
  As in Fig.~\ref{fig2}, but showing the spin correlation of protons $1'3'$,
  $\langle \sigma_y^{1'} \sigma_y^{3'} \rangle$.
}  
\label{fig3}
\end{center}
\end{figure}

\begin{figure}
\begin{center}
\begin{tabular}{c}
\resizebox{135mm}{!}{\includegraphics[angle=270]{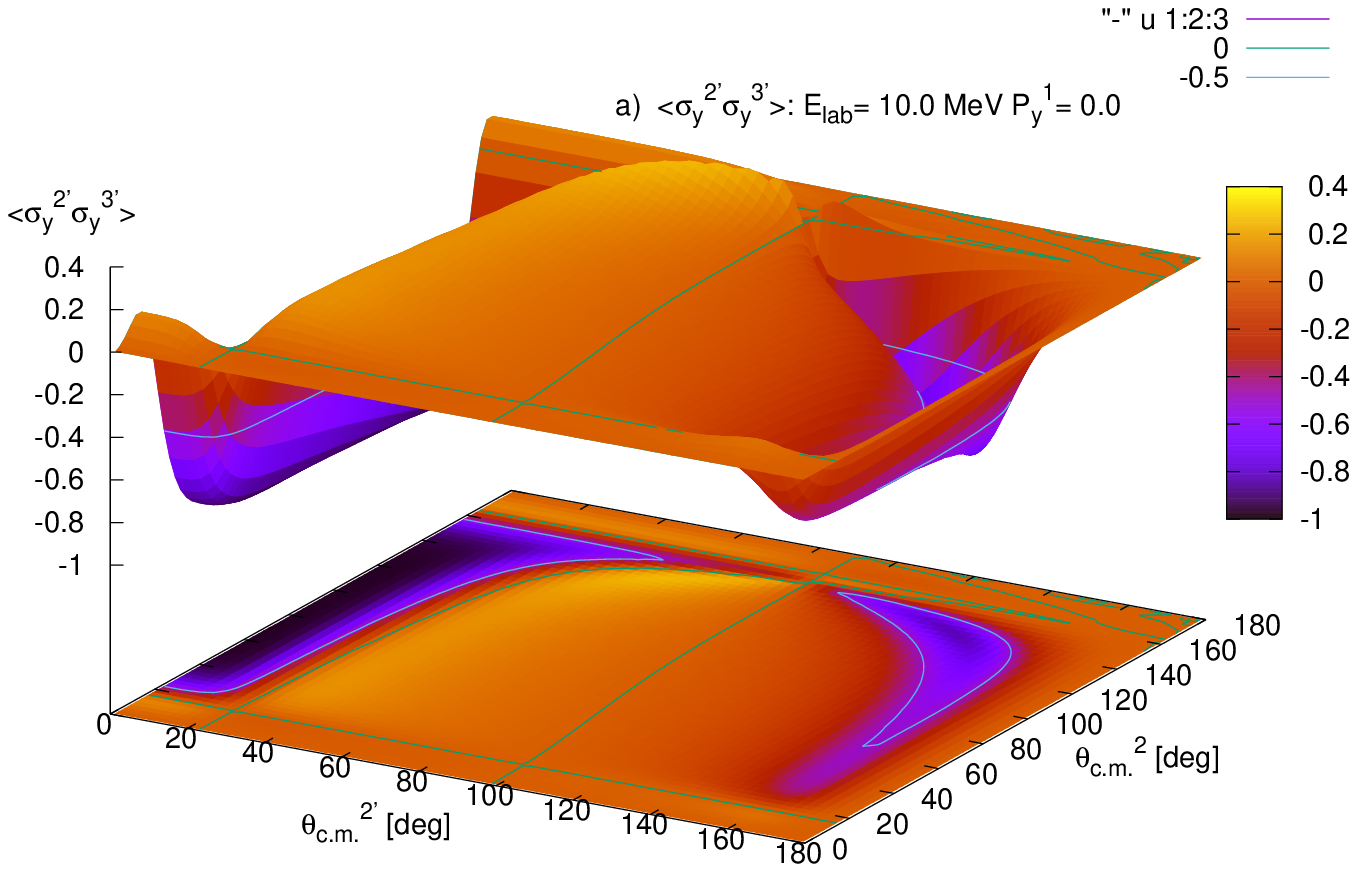}} \\
\resizebox{135mm}{!}{\includegraphics[angle=270]{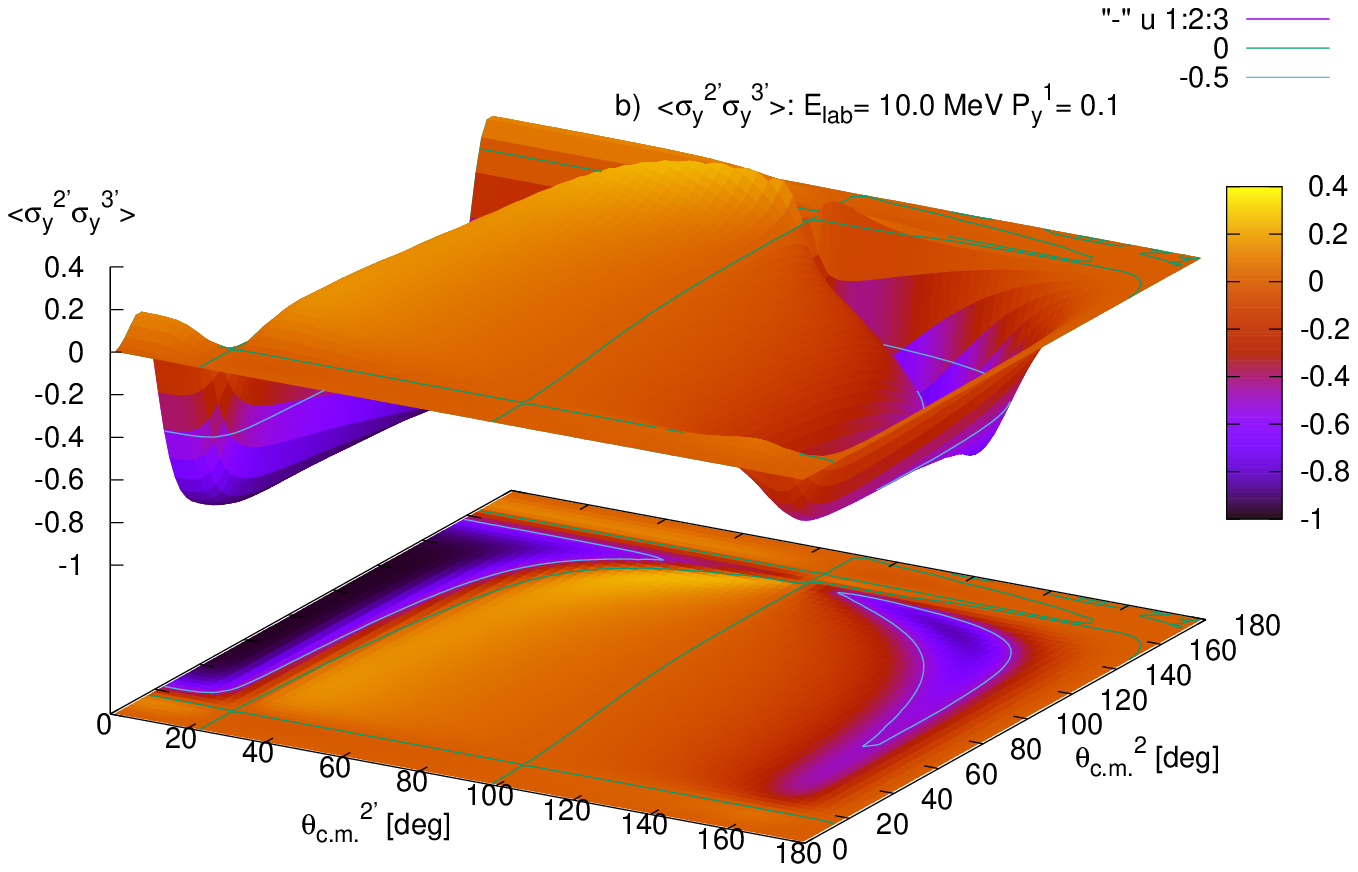}} \\
\end{tabular}%
\caption{
  (color online)
  As in Fig.~\ref{fig2}, but for the spin correlation of protons $2'3'$,
  $\langle \sigma_y^{2'} \sigma_y^{3'} \rangle$.
}  
\label{fig4}
\end{center}
\end{figure}

\begin{figure}
\begin{center}
\begin{tabular}{c}
\resizebox{135mm}{!}{\includegraphics[angle=270]{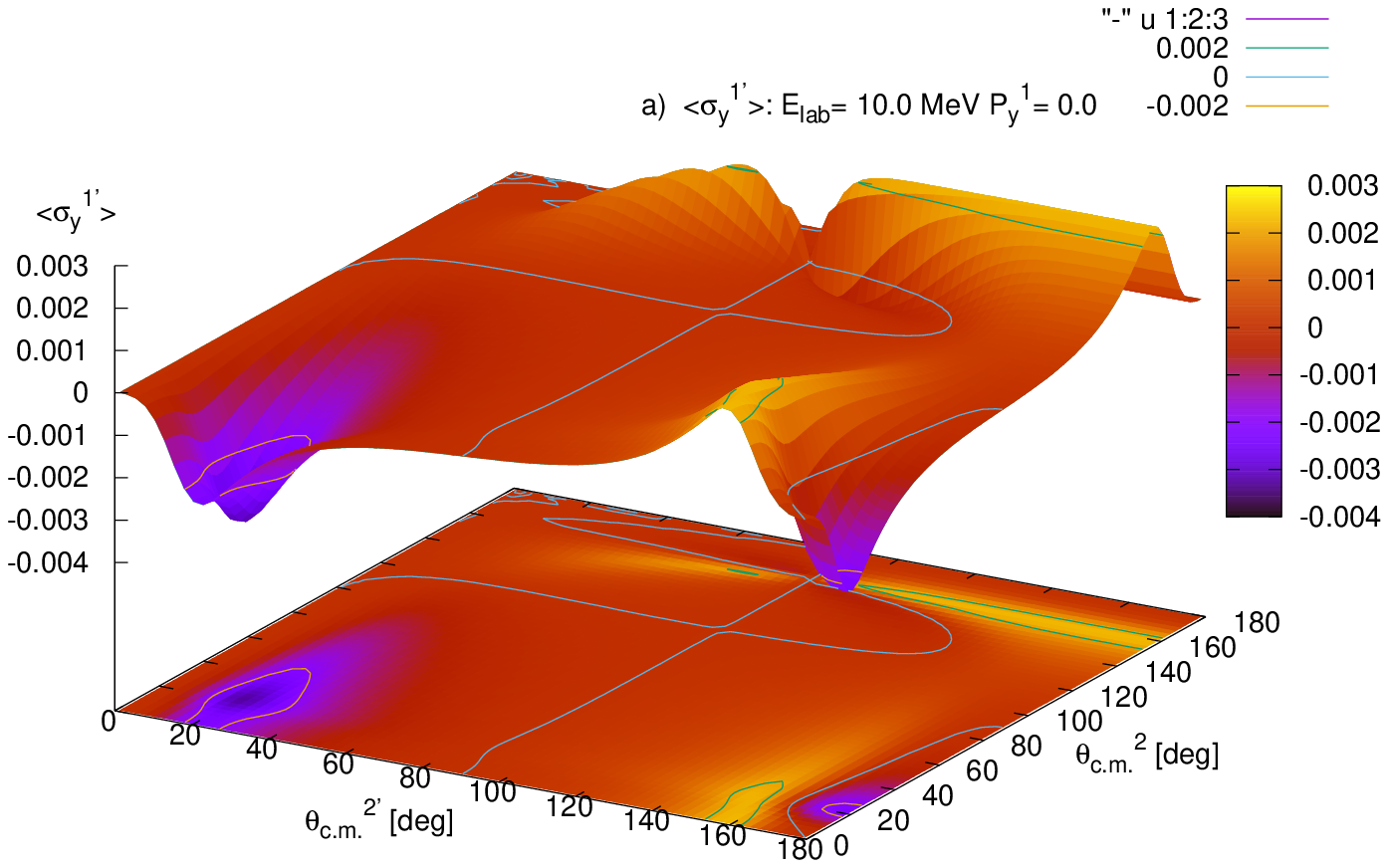}} \\
\resizebox{135mm}{!}{\includegraphics[angle=270]{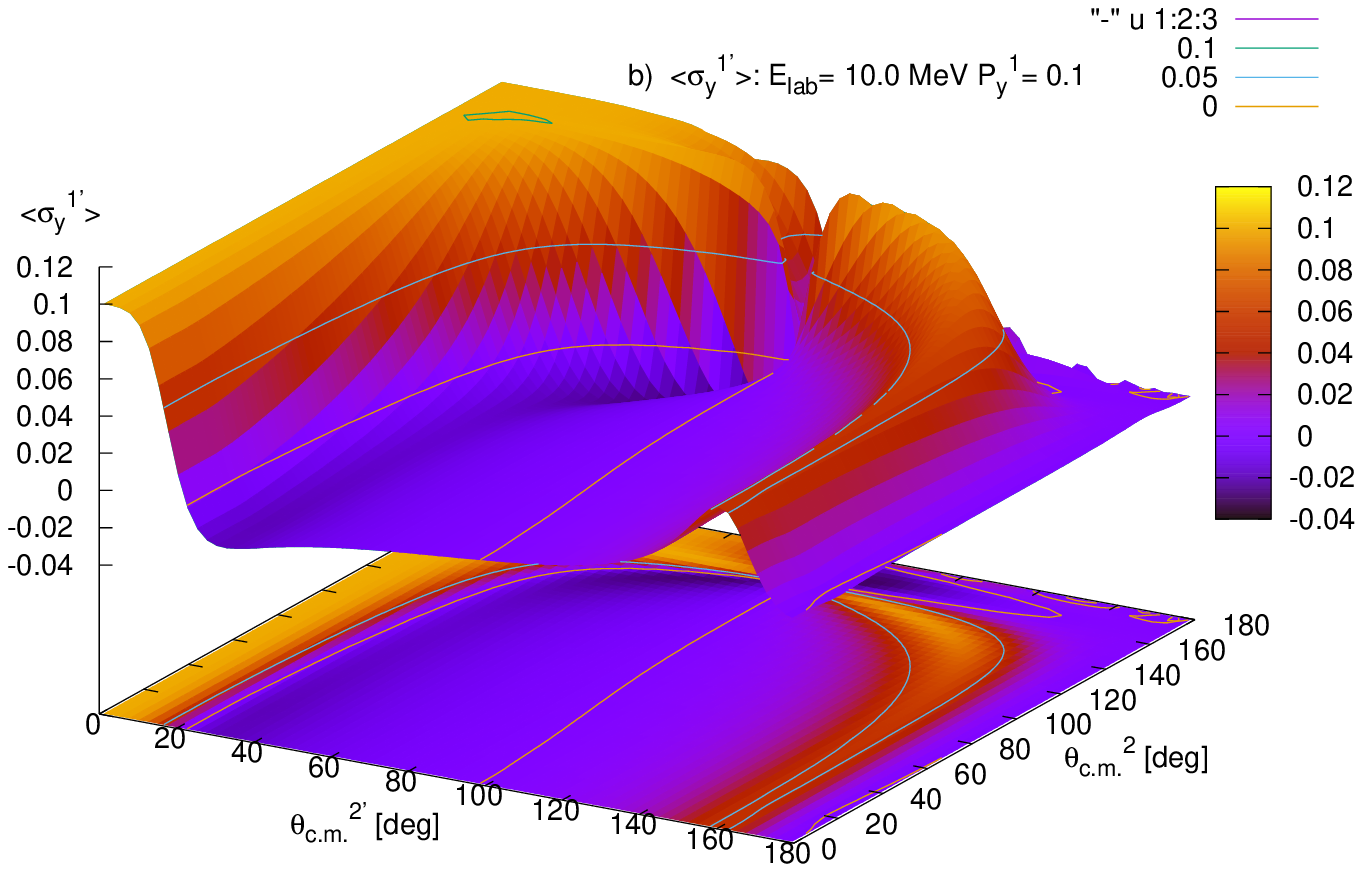}} \\
\end{tabular}%
\caption{
  (color online)
  As in Fig.~\ref{fig2}, but for the polarization of proton $1'$,
  $\langle \sigma_y^{1'} \rangle$.
}  
\label{fig5}
\end{center}
\end{figure}

\begin{figure}
\begin{center}
\begin{tabular}{c}
\resizebox{135mm}{!}{\includegraphics[angle=270]{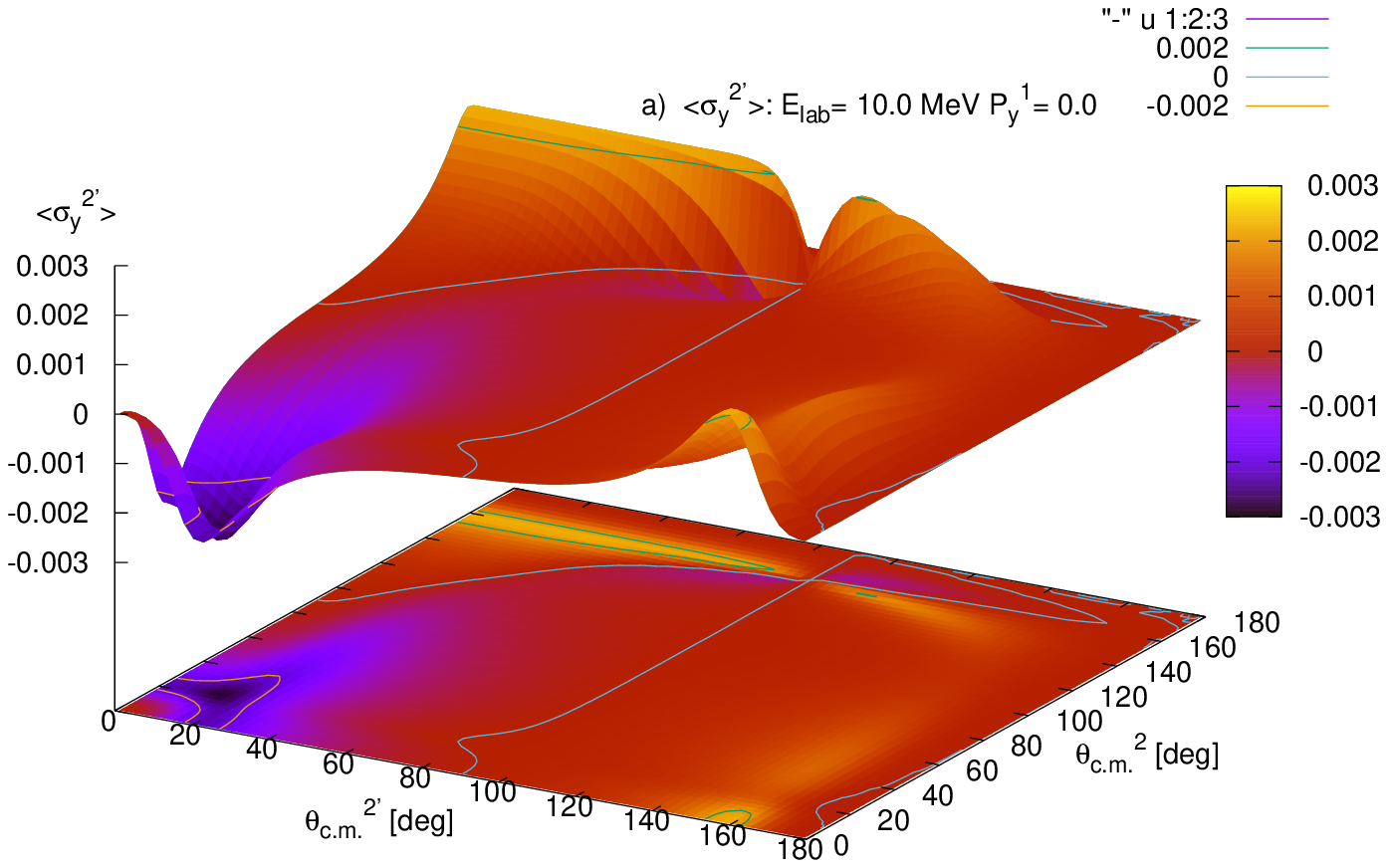}} \\
\resizebox{135mm}{!}{\includegraphics[angle=270]{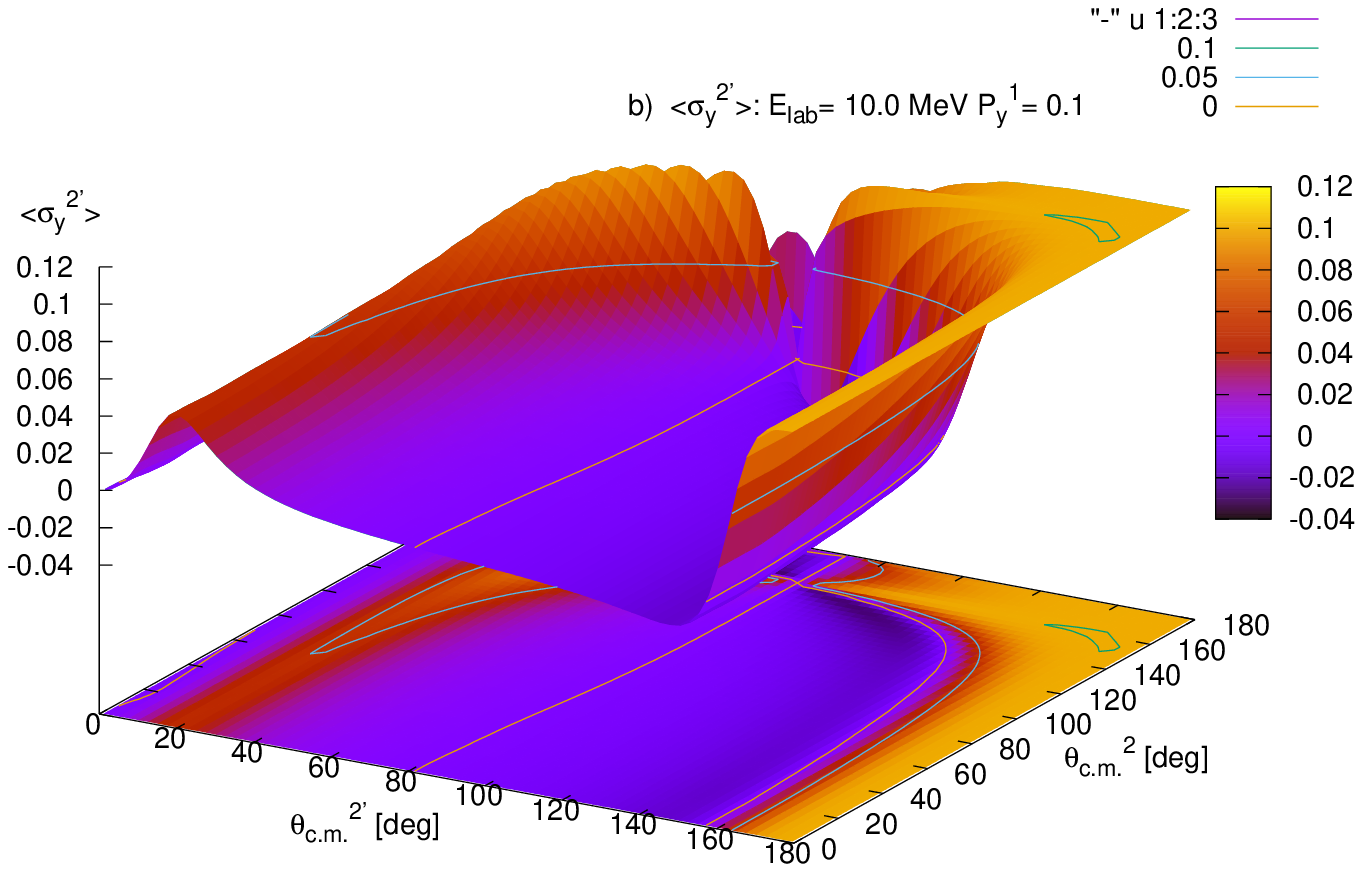}} \\
\end{tabular}%
\caption{
  (color online)
  As in Fig.~\ref{fig2}, but for the polarization of proton $2'$,
  $\langle \sigma_y^{2'} \rangle$.
}  
\label{fig6}
\end{center}
\end{figure}

\begin{figure}
\begin{center}
\begin{tabular}{c}
\resizebox{135mm}{!}{\includegraphics[angle=270]{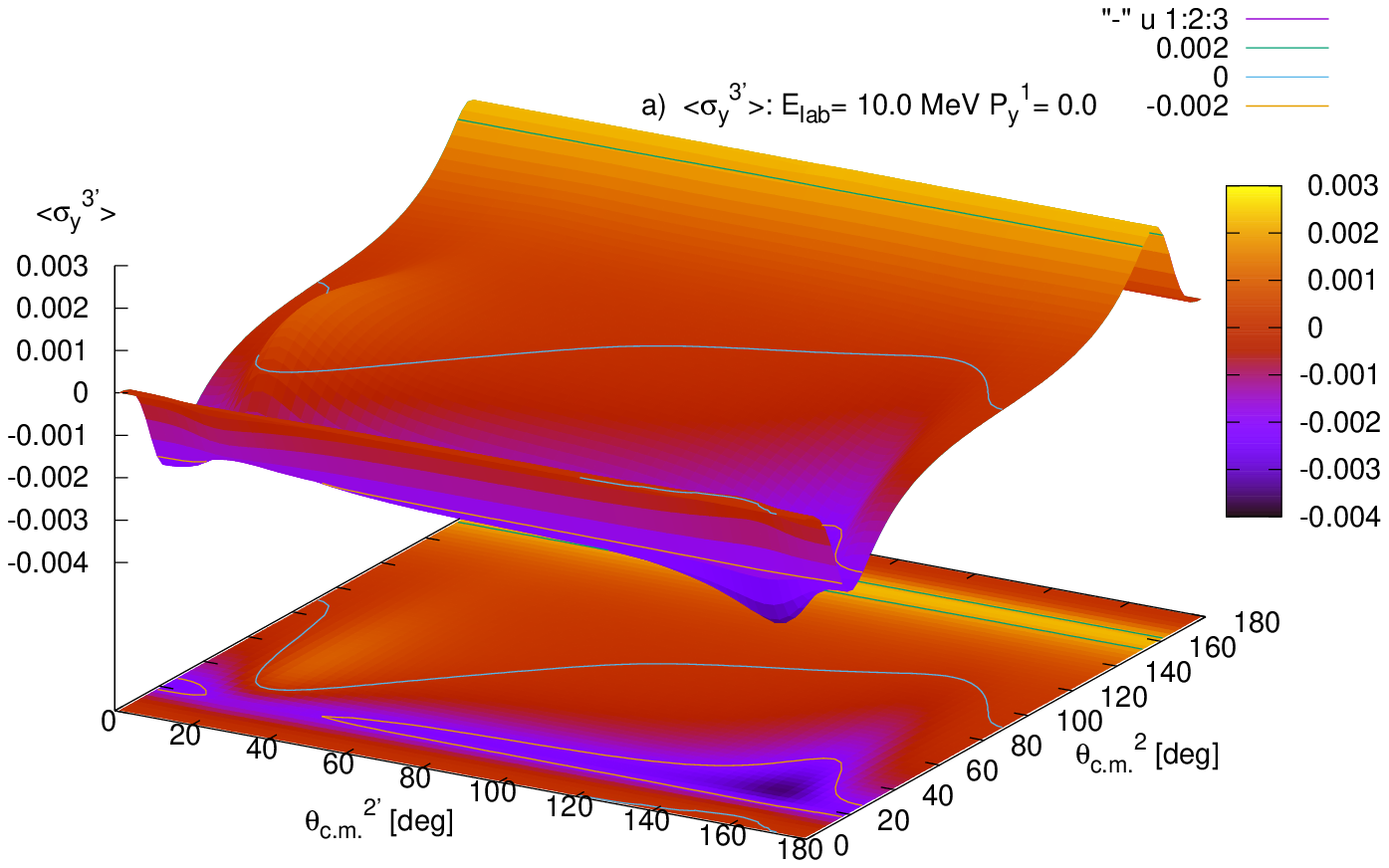}} \\
\resizebox{135mm}{!}{\includegraphics[angle=270]{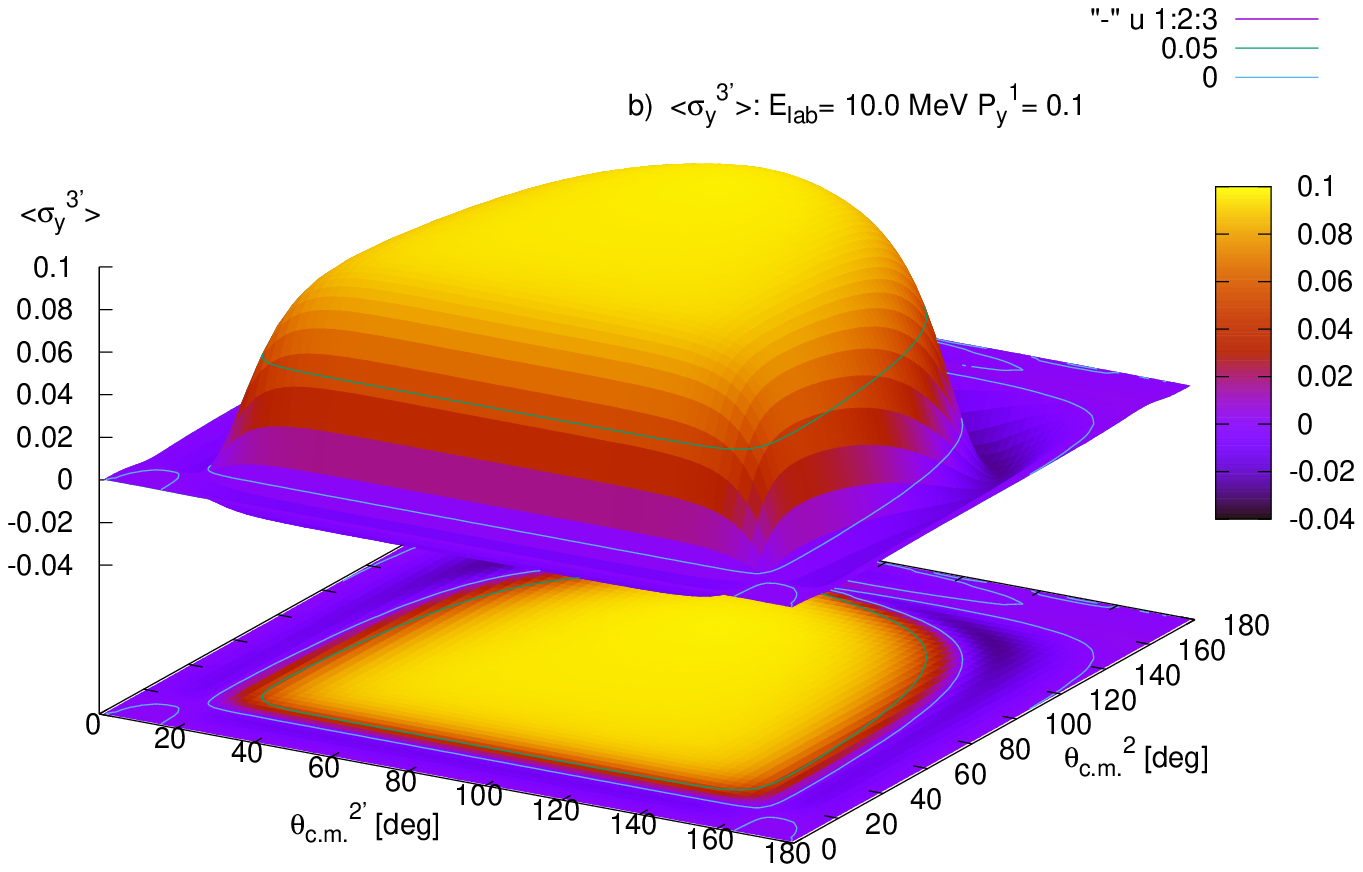}} \\
\end{tabular}%
\caption{
  (color online)
  As in Fig.~\ref{fig2}, but for the polarization of proton $3'$,
  $\langle \sigma_y^{3'} \rangle$.
}  
\label{fig7}
\end{center}
\end{figure}

\begin{figure}
\begin{center}
\begin{tabular}{c}
\resizebox{135mm}{!}{\includegraphics[angle=270]{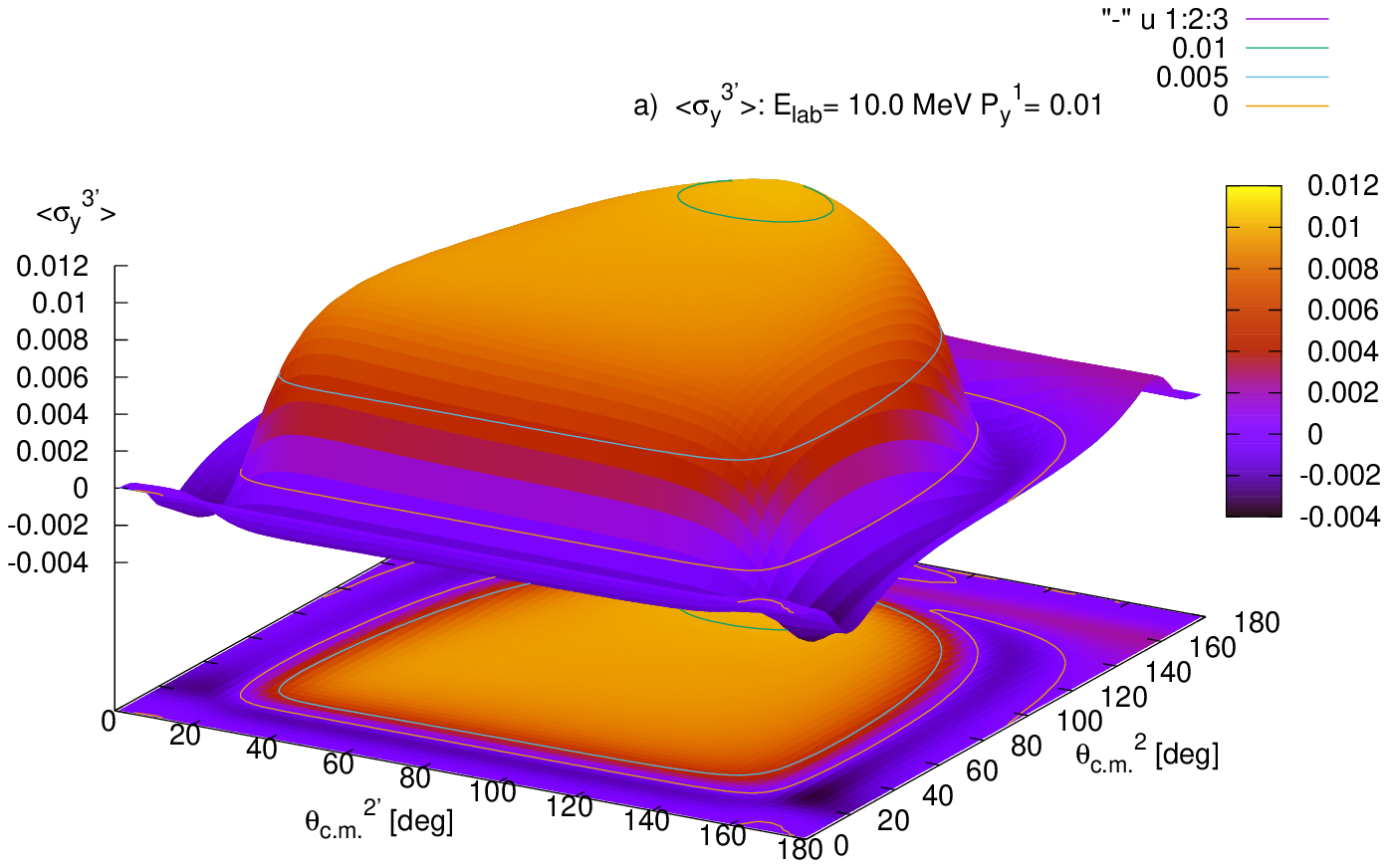}} \\
\resizebox{135mm}{!}{\includegraphics[angle=270]{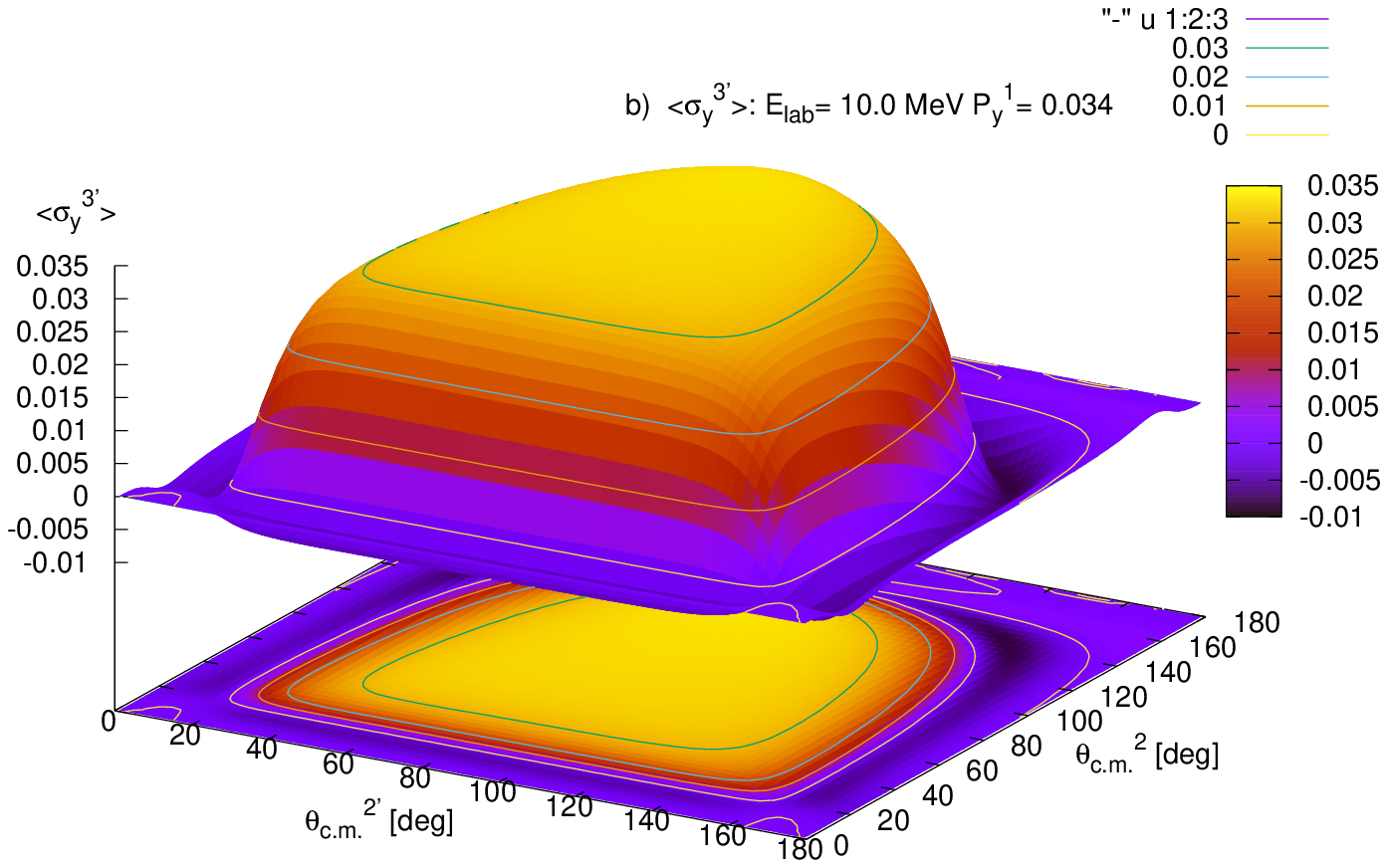}} \\
\end{tabular}%
\caption{
  (color online)
  As in Fig.~\ref{fig7}b, but for the polarization of the hydrogen
  target $1$, $P_y^{1} = 0.01$ (a) and $P_y^{1} = 0.034$ (b).
}  
\label{fig8}
\end{center}
\end{figure}

\begin{figure}
\includegraphics[scale=0.75]{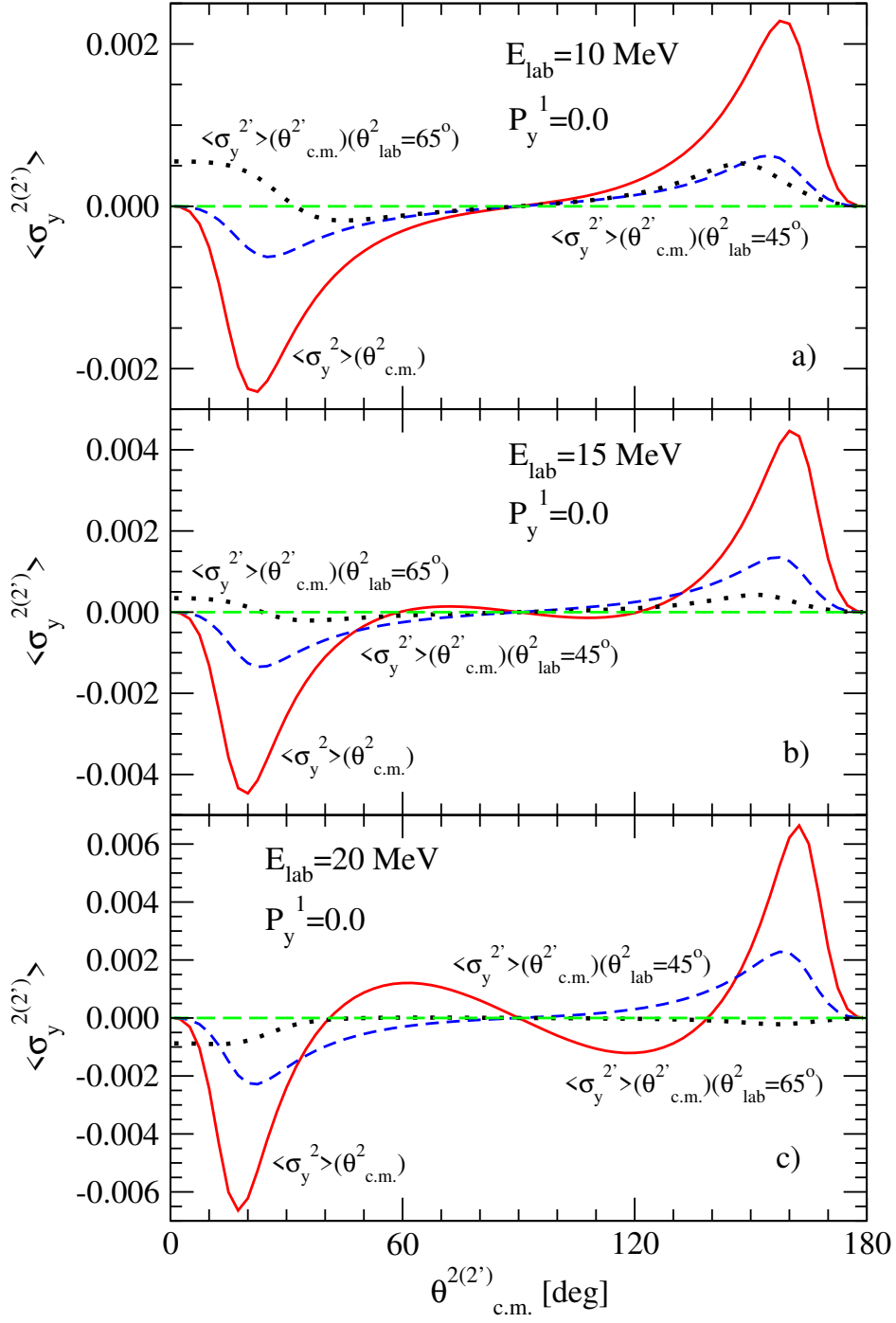}
\caption{
  (color online)
  Polarization $\langle \sigma_y^{2} \rangle$ of proton $2$ with the unpolarized
  hydrogen target $1$ removed (red solid line), shown as a function of
  the c.m. angle $\theta^{2}_{\mathrm{c.m.}}$. Dashed blue and dotted black
  lines: polarization $\langle \sigma_y^{2'} \rangle$ of proton $2'$
  after scattering of proton $2$ from the unpolarized hydrogen target $1$,
  for $\theta^{2}_{\mathrm{lab}} = 45^\circ$ and $65^\circ$, respectively,
  plotted versus $\theta^{2'}_{\mathrm{c.m.}}$.
  }
  \label{fig9}
\end{figure}

\begin{figure}
  \includegraphics[bb=31 217 410 520, scale=1.0]{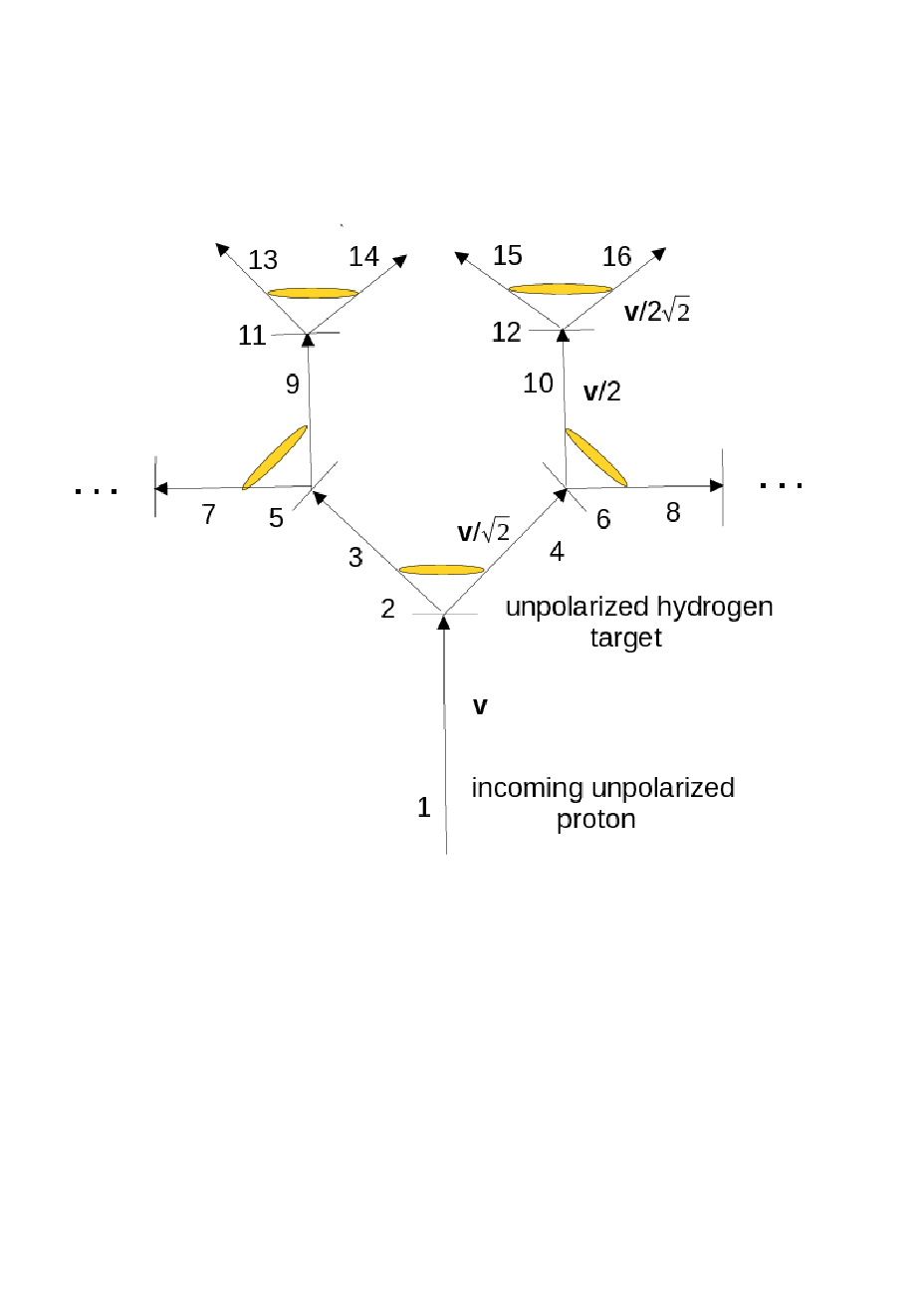}
\caption{
  (color online)
  Production of the entangled $pp$ pair $34$ in unpolarized $pp$ scattering,
  followed by subsequent scatterings of the outgoing protons on unpolarized
  hydrogen targets (indicated by lines perpendicular to the incident
  protons), leading to the formation of additional entangled $pp$ pairs.
  The orange ellipses denote entanglement within the $pp$ pairs. The
  velocities of the successive outgoing protons are reduced by
  a factor of $\sqrt{2}$.
  }
\label{fig1a}
\end{figure}

\begin{figure}
  \includegraphics[bb=19 385 424 605, scale=0.9]{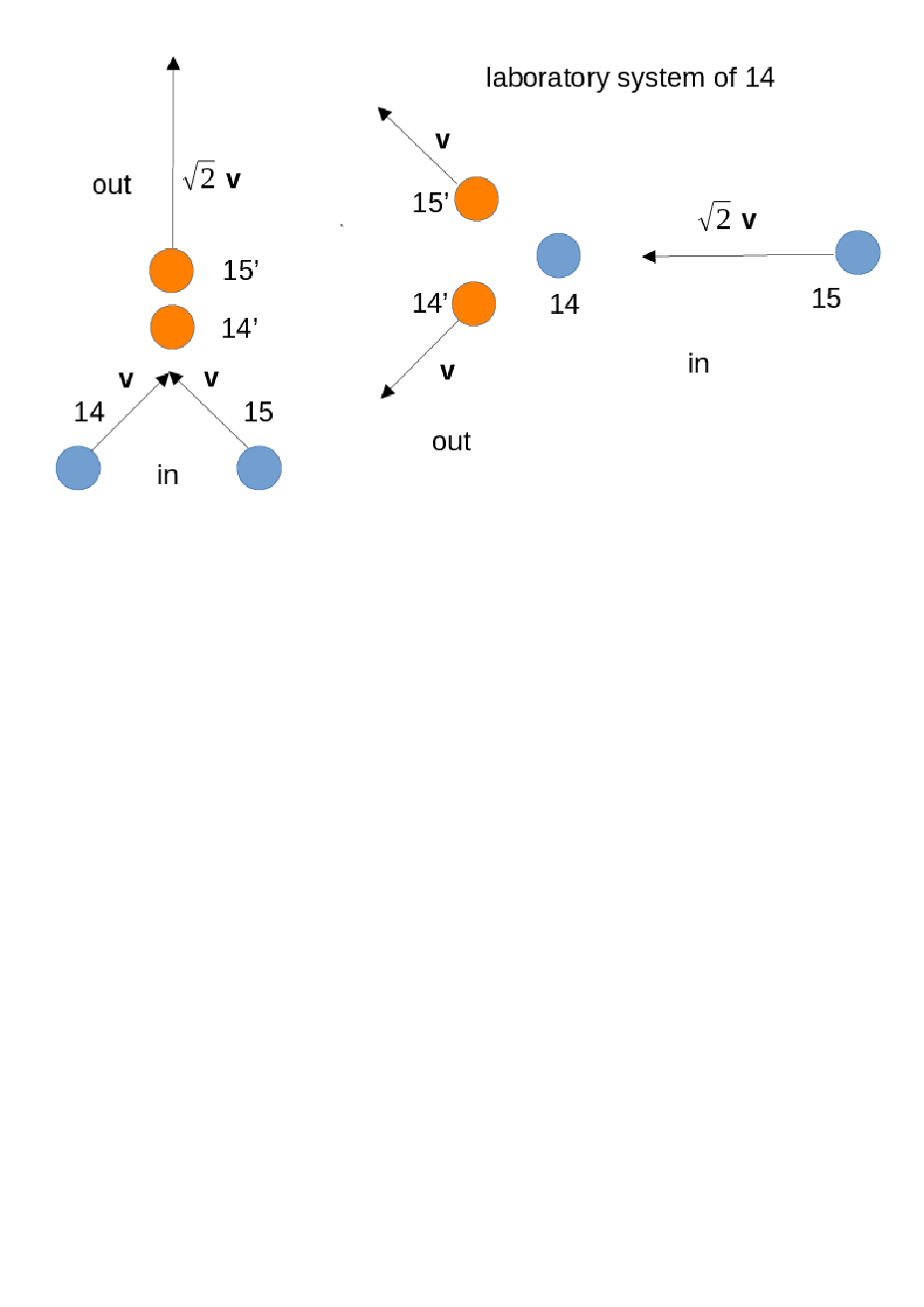}  
\caption{
  (color online)
  Kinematics of proton $14$ scattering from proton $15$ in the original
  coordinate system of Fig.~\ref{fig1a} and in the laboratory frame
  of proton $14$.
  }
\label{fig2a}
\end{figure}

Let the spin state of hydrogen target $1$ be characterized by a spin density
matrix with polarization vector $\vec P_1 = (0, P_y^1, 0)$. Consequently,
the initial spin density matrix of the three-proton system depicted
in Fig.~\ref{fig1} takes the form:
\begin{eqnarray}
  \rho_{\text{in}} = \rho_{\text{23}} \otimes \rho_{\text{1}} =
  \frac {1} {4}  M_{\text{23}} I^2 I^3 M_{\text{23}}^{\dagger}  \otimes
  \frac{1}{2} (I^1 + P_y^1 \sigma_y^1) 
   ~.
 \label{eq_ape1}
\end{eqnarray}

The final density matrix of the system, after proton $2$ has scattered
from hydrogen target $1$, is given by:
\begin{eqnarray}
  \rho_f &=& M_{12} \otimes I^3 ~  \rho_{\text{in}} ~   (M_{12} \otimes I^3)^{\dagger}
   ~.
 \label{eq_ape2}
\end{eqnarray}

The final polarization of proton $i'$, $\langle \sigma_y^{i'} \rangle$, 
 is therefore given by:
\begin{eqnarray}
  Tr (\rho_f ) \langle \sigma_y^{i'} \rangle &=&  
  Tr ( \rho_f^0 ) [ P_y^{i'~ind} +  P_y^1 K_y^{y'}(1 \to i') ]
   ~,
 \label{eq_ape3}
\end{eqnarray}
where  $P_y^{i'~ind}$ denotes the induced polarization of proton $i'$:
\begin{eqnarray}
  P_y^{i'~ind} &\equiv& \frac {Tr ( M_{12}
     M_{23}M_{23}^{\dagger} M_{12}^{\dagger} \sigma_y^{i'} ) }
  {Tr (M_{12}^{\dagger}M_{12}  M_{23}M_{23}^{\dagger} ) }
   ~,
 \label{eq_ape4}
\end{eqnarray}
and the polarization transfer coefficient  $K_y^{y'}(1 \to i')$   from
proton $1$ to proton $i'$, is given by:
\begin{eqnarray}
  K_y^{y'}(1 \to i') &=& \frac {Tr ( M_{12} M_{23} M_{23}^{\dagger}
    \sigma_y^{1} M_{12}^{\dagger} \sigma_y^{i'} ) }
  {Tr (M_{12}^{\dagger}M_{12}  M_{23}M_{23}^{\dagger} ) }
   ~.
 \label{eq_ape5}
\end{eqnarray}
Here, $\rho_f^0$  represents the final spin density matrix corresponding to
an unpolarized proton $1$ ($P_y^1=0$), and the trace of the final density
matrix is given by:
\begin{eqnarray}
  Tr (\rho_f )  &=&  \frac {1} {8} 
  [ Tr (M_{12} M_{23}M_{23}^{\dagger} M_{12}^{\dagger})
  +  P_y^1 Tr (M_{12} M_{23}M_{23}^{\dagger} \sigma_y^1  M_{12}^{\dagger} ) ]
   ~.
 \label{eq_ape3.1}
\end{eqnarray}

For the case when only one Bell term dominates the transition matrix $M$,
taking  $M = C | \psi^- \rangle  \langle \psi^- |$, which is approximately
valid at  $E_{lab} \approx 10$~MeV, one obtains for the matrix element
$\langle m_1 m_2 \vert  M \vert m'_1 m'_2 \rangle$:
\begin{eqnarray}
  \langle m_1 m_2 \vert  M \vert m'_1 m'_2 \rangle &=& \frac {C} {2}
  ( \delta_{m_1+} \delta_{m'_1+} \delta_{m_2-} \delta_{m'_2-}
  - \delta_{m_1+} \delta_{m'_1-} \delta_{m_2-} \delta_{m'_2+} \cr
&& - \delta_{m_1-} \delta_{m'_1+} \delta_{m_2+} \delta_{m'_2-}
 + \delta_{m_1-} \delta_{m'_1-} \delta_{m_2+} \delta_{m'_2+} )
  ~.
\label{eq_ape6}
\end{eqnarray}
In the following, we use the shorthand notation $+(-)$ for $+(-)\frac{1}{2}$ 
 to denote the magnetic quantum numbers. 

 An identical expression is obtained for the matrix element  $MM^\dagger$ upon
 replacing $C$ with $\vert C \vert^2$.

 If either the first or the second scattering is dominated by such a
 contribution and
$$
M M^{\dagger} = |C|^2 |\psi^- \rangle \langle \psi^- |,
$$
the traces entering Eqs.~(\ref{eq_ape4}), (\ref{eq_ape5}),
and (\ref{eq_ape3.1}) reduce to:
\begin{eqnarray}
  Tr (M_{12}^{\dagger}M_{12}  M_{23}M_{23}^{\dagger} ) &=&
  \frac {\vert C \vert^2}  {2} 
        [  Tr ( M_{12} M_{12}^{\dagger} ) \cr
      &-&  \sum_{m1} 
   ( \langle m_1 - \vert  M_{12}^{\dagger}M_{12} \vert m_1 + \rangle 
          +\langle m_1 + \vert  M_{12}^{\dagger}M_{12} \vert m_1 - \rangle ]
 \label{eq_ape7.1}        
\\
Tr ( M_{12} M_{23}M_{23}^{\dagger} M_{12}^{\dagger} \sigma_y^{3'})  &=&
\frac {\vert C \vert^2}  {2} Tr ( M_{23} M_{23}^{\dagger}  \sigma_y^{3'})
 \label{eq_ape7.2} 
\\
Tr ( M_{12} M_{23} M_{23}^{\dagger}
    \sigma_y^1 M_{12}^{\dagger} \sigma_y^{3'} ) &=&
- \frac {\vert C \vert^2}  {2}
( - \langle + + \vert  M_{12}^{\dagger}M_{12} \vert - - \rangle 
   +\langle + - \vert  M_{12}^{\dagger}M_{12} \vert - + \rangle \cr
 &&~~~~~  + \langle - + \vert  M_{12}^{\dagger}M_{12} \vert + - \rangle
   - \langle - - \vert  M_{12}^{\dagger}M_{12} \vert + + \rangle )
 \label{eq_ape7.3}
   \\
Tr ( M_{12} M_{23}M_{23}^{\dagger} M_{12}^{\dagger} \sigma_y^{1'(2')})  &=&
\frac {\vert C \vert^2}  {2} Tr ( M_{12} M_{12}^{\dagger}  \sigma_y^{1'(2')} ) 
 \label{eq_ape7.4}
\\
Tr ( M_{12} M_{23} M_{23}^{\dagger}
\sigma_y^1 M_{12}^{\dagger} \sigma_y^{1'(2')} ) &=& 
\frac {\vert C \vert^2}  {2}
Tr ( \sigma_y^1 M_{12}^{\dagger} \sigma_y^{1'(2')} M_{12}) 
 \label{eq_ape7.5}
\\
Tr (M_{12} M_{23}M_{23}^{\dagger}  \sigma_y^{1} M_{12}^{\dagger} ) &=&
\frac {\vert C \vert^2}  {2} Tr ( M_{12}^{\dagger} M_{12} \sigma_y^{1} )
\label{eq_ape7.6}
~.
\end{eqnarray}
The last trace is needed for the calculation of
$Tr(\rho_f)$ of Eq.~(\ref{eq_ape3.1}).

When in the first and second scattering one Bell term
dominates, 
the calculation of the traces in (\ref{eq_ape7.1})–(\ref{eq_ape7.6}) can
be carried out and the observables can be evaluated.

Such domination in both transition matrices $M_{12}$ and $M_{23}$  
occurs at low energies $E_{lab} \approx 10$~MeV to a very good
approximation in the angular region
$\theta_{c.m.}^2$ and  $\theta_{c.m.}^{2'} \in (55^\circ,125^\circ)$
\cite{wit_unp_pd}. 
At these angles, the sum over nondiagonal terms in the trace of
Eq.~(\ref{eq_ape7.1}) vanishes, and:
\begin{eqnarray}
Tr (M_{12}^{\dagger}M_{12}  M_{23}M_{23}^{\dagger} ) &=&
\frac {\vert C \vert^2}  {2} Tr( M_{12}^{\dagger}M_{12} ) =
\frac {\vert C \vert^2}  {2} Tr( M_{23}^{\dagger}M_{23} )
\label{eq_ape7.7}
  ~.
\end{eqnarray}

Consequently, the induced contribution $P_y^{1'(2')~ind}$ to the final
polarization of protons $1'$ and $2'$ in Eq.~(\ref{eq_ape4}) is equal to
the induced polarization in the second scattering of proton $2$ from proton
$1$ (see Eqs.~(\ref{eq_ape7.4}) and (\ref{eq_ape7.7}), where
only the transition matrix $M_{12}$
enters the expression) (see also Refs.~\cite{ohlsen1972,wit_spin_np_entangl}).

Similarly, the contribution to the final polarization arising from
the polarization transfer coefficients  $K_y^{y'}(1 \to 1'(2'))$  is, for
both protons $1'$ and $2'$, given by the polarization transfer coefficient
of the second $pp$ scattering 
(see Eqs.~(\ref{eq_ape7.5}) and (\ref{eq_ape7.7})).

In both cases, the corresponding observables are those of the second $pp$
scattering, evaluated at the energy of the incident proton $2$ 
at the scattering angle $\theta_{c.m.}^2$

For proton $3'$, the induced contribution to its final polarization in
Eq.~(\ref{eq_ape3}), $P_y^{3'~\mathrm{ind}}$, given by Eqs.~(\ref{eq_ape7.2})
and (\ref{eq_ape7.7}), is determined, in the angular region of strong
entanglement considered here, by the induced polarization in the first
$pp$ scattering, since only the transition matrix $M_{23}$
enters Eqs.~(\ref{eq_ape7.2}) and (\ref{eq_ape7.7}).

In this angular region, the polarization-transfer coefficient satisfies
$K_y^{y'}(1 \to 3') = 1$ (see Eqs.~(\ref{eq_ape7.3}) and (\ref{eq_ape7.7}),
together with
$M_{12}^{\dagger}M_{12} = |C|^2 |\psi^-\rangle_{12} ~_{12}\langle\psi^-|$).
Furthermore,
$$
\mathrm{Tr}\left(M_{12} M_{23} M_{23}^{\dagger}\sigma_y^1 M_{12}^{\dagger}\right)=0
$$
(Eq.~(\ref{eq_ape7.6})), which implies
$\mathrm{Tr}(\rho_f)=\mathrm{Tr}(\rho_f^0)$.
Consequently, the final polarization of proton $3'$ is given by
$$
\langle \sigma_y^{3'} \rangle = P_y^{3'~\mathrm{ind}} + P_y^1.
$$

It is evident that the value $K_y^{y'}(1 \to 3') = 1$ enables the
teleportation-like transfer of polarization and can be directly
associated with the dominance of a single Bell component in the
transition matrix  $M_{12}$ of the second scattering of proton $2$
from proton $1$ (see Eq.~(\ref{eq_ape7.3})).

We now consider the final spin correlations between different protons
following the scattering of proton $2$ from proton $1$. They are given by:
\begin{eqnarray}
  Tr (\rho_f ) \langle \sigma_y^{i'} \sigma_y^{j'} \rangle &=&  
  Tr ( \rho_f^0 ) [ \langle \sigma_y^{i'} \sigma_y^{j'} \rangle^{ind} +
    P_y^1 K_{0y}^{y'y'}(1 \to i'j') ] 
   ~.
 \label{eq_ape8}
\end{eqnarray}
As in the case of the final polarizations, the correlations consist of
two contributions. The first is independent of the polarization of
proton $1$ and is referred to as the induced correlation
$\langle \sigma_y^{i'} \sigma_y^{j'} \rangle^{ind}$  for the $i'j'$  pair:
\begin{eqnarray}
  \langle \sigma_y^{i'} \sigma_y^{j'} \rangle^{ind} &\equiv& \frac {Tr (M_{12}
     M_{23}M_{23}^{\dagger} M_{12}^{\dagger} \sigma_y^{i'} \sigma_y^{j'} ) }
  {Tr (M_{12}^{\dagger}M_{12}  M_{23}M_{23}^{\dagger} ) }
   ~,
 \label{eq_ape9}
\end{eqnarray}
and the second is the contribution to the final correlation induced by 
the polarization of proton $1$, $P_y^1 K_{0y}^{y'y'}(1 \to i'j')$ with:
\begin{eqnarray}
  K_{0y}^{y'y'}(1 \to i'j') &=& \frac {Tr ( M_{12} M_{23} M_{23}^{\dagger}
    \sigma_y^{1} M_{12}^{\dagger} \sigma_y^{i'} \sigma_y^{j'}) }
  {Tr (M_{12}^{\dagger}M_{12}  M_{23}M_{23}^{\dagger} ) }
   ~.
 \label{eq_ape10}
\end{eqnarray}
We adopt a convention analogous to that introduced in
Ref.~\cite{wit_spin_np_entangl}, where the corresponding quantity
 $K_{0y}^{y'y'}(1 \to i'j')$ was
referred to as the single-spin correlation transfer coefficient.

Assuming that the matrix $M_{23}$ is dominated by the single term
$|\psi^- \rangle \langle \psi^- |$, one obtains
\begin{eqnarray}
  Tr (M_{12}M_{23}M_{23}^{\dagger} M_{12}^{\dagger} \sigma_y^{i'} \sigma_y^{j'} )  &=&
\frac {\vert C \vert^2} {2} Tr(M_{12} M_{12}^{\dagger} \sigma_y^{i'} \sigma_y^{j'} ) 
~,
\label{eq_ape11}
\\
  Tr (M_{12}M_{23}M_{23}^{\dagger} \sigma_y^{1} M_{12}^{\dagger}
  \sigma_y^{i'} \sigma_y^{j'} )  &=&
  \frac {\vert C \vert^2} {2} Tr(M_{12} \sigma_y^{1}
  M_{12}^{\dagger} \sigma_y^{i'} \sigma_y^{j'} ) 
   ~,
 \label{eq_ape11.1}
\end{eqnarray}
and, consequently, for the induced spin correlation
$\langle \sigma_y^{i'} \sigma_y^{j'} \rangle^{\mathrm{ind}}$ one finds
\begin{eqnarray}
  \langle \sigma_y^{i'} \sigma_y^{j'} \rangle^{ind} &=& \frac {Tr (M_{12}
   M_{12}^{\dagger} \sigma_y^{i'} \sigma_y^{j'} ) }
  {Tr (M_{12}^{\dagger} M_{12} ) }
   ~,
 \label{eq_ape12}
\end{eqnarray}
whereas the single-spin correlation transfer coefficient
$K_{0y}^{y'y'}(1 \to i'j')$ is given by
\begin{eqnarray}
  K_{0y}^{y'y'}(1 \to i'j') &=& \frac {Tr (M_{12} \sigma_y^{1} 
   M_{12}^{\dagger} \sigma_y^{i'} \sigma_y^{j'} ) }
  {Tr (M_{12}^{\dagger} M_{12} ) }
   ~. 
 \label{eq_ape12.1}
\end{eqnarray}
Both quantities are identical to the corresponding observables in $pp$
scattering, with proton $2$ scattering from proton $1$.

Assuming, in addition, that the matrix $M_{12}$ is also dominated by
the single term
$|\psi^- \rangle \langle \psi^- |$, one obtains
$$\langle \sigma_y^{1'} \sigma_y^{2'} \rangle^{\mathrm{ind}} = -1,$$ 
whereas
$\langle \sigma_y^{1'} \sigma_y^{3'} \rangle^{\mathrm{ind}}$,
$\langle \sigma_y^{2'} \sigma_y^{3'} \rangle^{\mathrm{ind}}$,
as well as all
$K_{0y}^{y'y'}(1 \to i'j')$, vanish.

It is evident that the dominance of a single Bell-state term in the
transition matrix $M_{12}$ is responsible for the formation of the
strongly entangled proton pair $1'2'$, just as the analogous dominance in
$M_{23}$ led to the formation of the pair $23$ in the first scattering.
The formation processes of these entangled pairs are completely independent,
and no teleportation of entanglement from the pair $23$ to the pair
$1'2'$ takes place, contrary to what was erroneously suggested in
Ref.~\cite{wit_unp_pd}.

It should be emphasized that these two entangled states are not
identical, but differ
in the entanglement-degrading contributions associated with the
nonvanishing polarizations of the constituent protons. For the pair $23$,
these are equal to the induced polarizations generated by the
transition matrix $M_{23}$, whereas for the  pair $1'2'$ it is $M_{12}$ that
yields different induced polarizations, since the energy of the incoming
proton $2$ differs from that of the incoming unpolarized proton in the
first scattering.



\acknowledgments

This work was supported by the National Science Centre,
Poland under Grant
IMPRESS-U 2024/06/Y/ST2/00135.   
The numerical calculations were partly performed on the supercomputers of
the JSC, J\"ulich, Germany.




%


\clearpage

\end{document}